# The thermodynamic roots of synergy and its impact on society

Klaus Jaffe

**Information and Energy are the Gods of Genesis** [1]

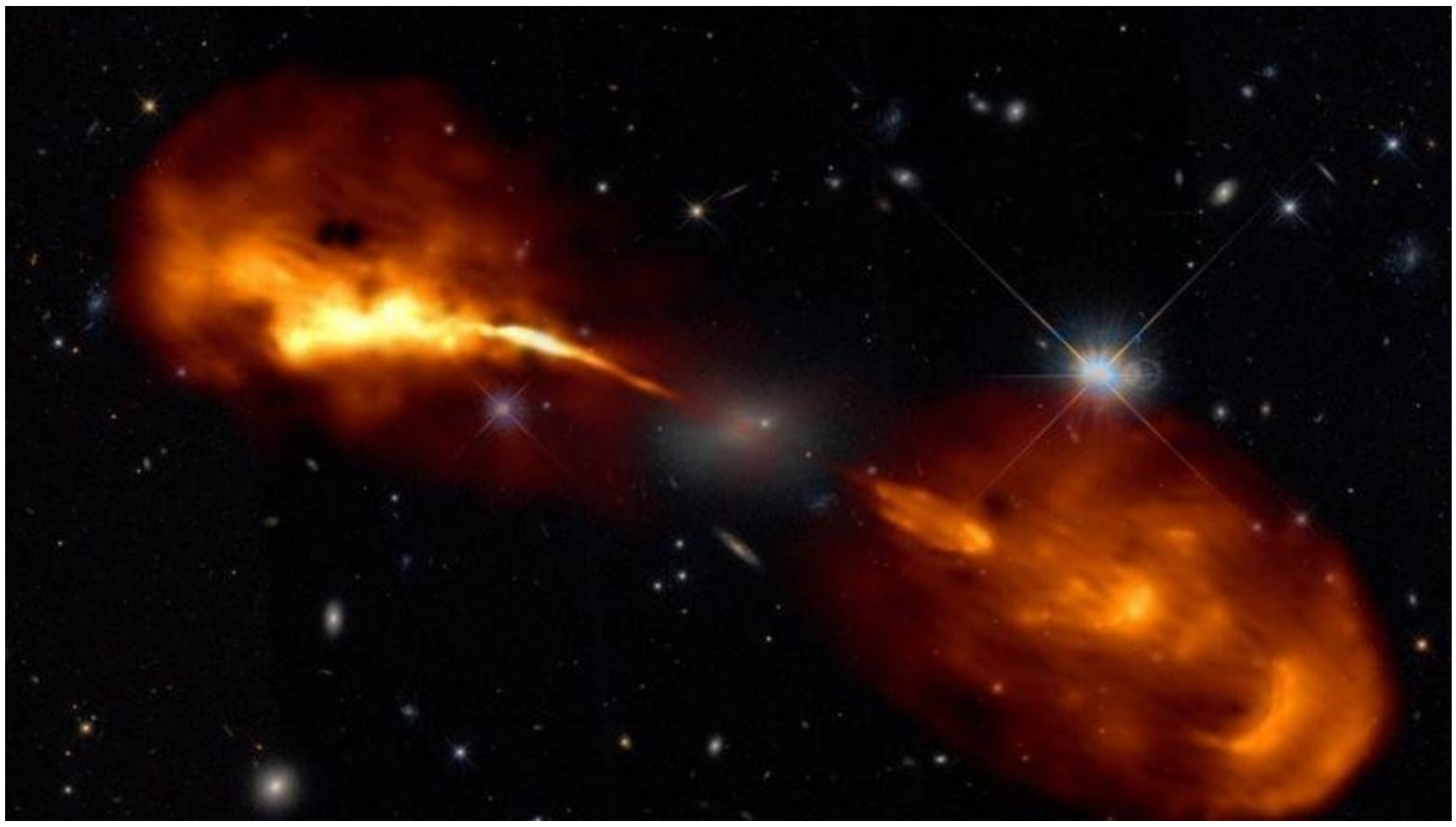

*Photograph of the most powerful release of free energy sighted in our universe, possible due to synergistic interactions of two massive black holes.*

*Researchers at Durham University in the UK combined radio signals from over 70,000 antennae across nine European countries to capture this image in 2021*

*Most forces in nature cause destruction, competition and conflict. A few processes, such as life, strive in the opposite direction, evolving ever more sophisticated and complex organisms despite being surrounded by death and chaos. Societies evolve, creating new structures and destroying old ones. Cooperation and synergy drive these biophilic processes that are intricately woven together through flows of information and energy. But not any information is relevant in deriving synergistic cooperation. Too much noise in the information exchange and the system collapses destroying the agents sustaining it. Novel insights in irreversible thermodynamics of complex systems help us to better understand these processes. This allows us to consiliently integrate evolutionary dynamics in diverse fields of knowledge including physics, chemistry, biology, social sciences, ethics and the arts. The theoretical and practical consequences of this new synthesis are still not completely fathomable but are already impressive. They open a window into a new world that bridges economics and biology; politics and empirical sciences; law and physics; sociobiology and thermodynamics; ethics and arts. The underlying wisdom is that free energy and useful work is only possible with the right type of order and information. When this happens, synergies emerge creating novel possibilities and new futures for ourselves and our society.*

## 1. Introduction

The usefulness of science in tackling social and economic problems has been limited. This is a reflection of the immaturity of scientific disciplines dealing with these complex questions. An example is our understanding of a very simple everyday phenomena in economics: “Money”. The latest attempt to explain money is the “Modern Monetary Theory”, which proposes novel twists of the nature of money and the working of economies with important implications in politics. This view, however, is still very incomplete. It is not “consilient”, or in agreement, with other areas of science whose theories are well established and tested, such as thermodynamics. In developing more consilient economic and social theories, we must not contradict general natural principles that are at the base of these phenomena. This book explores a more comprehensive scientific understanding of the roots of social dynamics with a focus on synergy.

One problem in understanding multidimensional phenomena is its complexity. Our brains, even with the help of advanced computers, have limits in grasping the fine net of non-linear interactions underlying the phenomena that produce synergy. Thus, searching analytical solutions to complex economic problems is irrational. For example, the number of possible behavioral and economic interactions between humans on our planet is larger than the number of atoms in the universe. No finite mathematical solution can be expected from a detailed analytic formulation of these interactions. However, nature has shown us the way to handle such seemingly irrational problems: It invented Biological Evolution.

Biological Evolution solved the problem of handling complex interactions of matter and energy between living organisms. As the possible mix of random processes, diversity variations, mutations and combinations of genes and organisms is unfathomably large, no rational solution of a precise analytical solution to this problem exists. Biological evolution solved this problem by riding on serendipity, selecting genes and organisms through the differential survival rates of these agents. In doing so, it favors agents and combinations of agents that improve the synergy of the interacting parts. That is, Nature has no advanced plan to produce better adapted and more sophisticated organisms; it uses serendipity and natural selection to do the job. This selection can act directly on the organism, or indirectly through the web of interactions that might confer advantages or limitations to it. This indirect selection mechanism is called “Inclusive Fitness” and its extension to economics and other human social phenomena is called “Extended Inclusive Fitness”. The unraveling of these dynamics has allowed modern science to build simulation models that mimic these complex natural social interactions and apply them to specific economic and social processes.

The knowledge we obtained from these simulations include many basic facts. One is that in all evolutionary processes in biological, cultural and physical (as used in artificial intelligence) settings, three conditions must apply: liberty, diversity and tradition.

- **Liberty** means that agents must be free to interact with others in many different ways and the action of each individual affects the well-being of others. No agents are insignificant, although some might be more influential than others.

- **Diversity** means that no single agent is identical to another. Large variations are present and

continuously novel random variations appear.

- **Tradition** means that there must be a form of cultural and or biological learning or of transmission of information from one generation to another, or one agent to another, such as heredity or culture.

In these processes, optimization of the agents and its interactions depends on location, leading evolution to produce a variety of forms, each adapted to a given set of constraints. Rarely only one solution to a complex problem emerges through evolution, although there might exist homologies and analogies.

Here we explore one aspect of this evolutionary dynamics, the emergence of synergies, or of interactions that benefit all or most of the interacting parts. As these synergies can not be predicted beforehand, only empirical results can uncover them. Consequently, the study of synergy is a novel empirical natural science, considered as an extension to thermodynamics.

This book is part of a series that follows a logical string of ideas. The first book: **What is Science: An**

**interdisciplinary evolutionary view**, makes a heuristics analysis, concluding that empirical science is the only way to accumulate knowledge of our world that allows us to approach reality with ever greater precision. The second book, **The Wealth of Nations: Complexity Science for an Interdisciplinary Approach in Economics**, explores empirical methods to advance our understanding of economics and society. This third book of the series exploring **Synergy**, expands this heuristics to all sorts of social phenomena, including human society and business management.

**While reading this book just skip over parts you do not understand or are not interested in. This world has many windows and there are many ways to grasp Synergy.**

# 2. What is Synergy

## 2.1 Definitions

The available dictionaries give us the following definitions for Synergy:

- The interaction of two or more agents or forces so that their combined effect is greater than the sum of the individual effects. (The Free Dictionary)

- A state in which two or more things work together in a particularly fruitful way that produces an effect greater than the sum of their individual effects. Expressed also as "the whole is greater than the sum of its parts. (Business Dictionary)

- The interaction or cooperation of two or more organizations, substances, or other agents to produce a combined effect greater than the sum of their separate effects. ‘the synergy between artist and record company’ (Oxford Dictionaries)

- The creation of a whole that is greater than the simple sum of its parts. The term synergy comes from the Attic Greek word συνεργία “synergia” from synergos, συνεργός, meaning "working together". (Wikipedia)

These definitions describe synergy as a process that is recognized by its result, but say nothing about how to achieve it nor what the roots of synergy are. How can something be more than the sum of its parts? If synergy relates to the phenomena where the output of a system is not explained by the simple sum of the output of each part: How do some interactions manage to produce synergy and others not?

The term synergy in science was used in neuromuscular physiology by Charles Scot Sherrington (born in 1857 in Islington; died 1952 in Eastbourne) when he described the integrative action of the nervous system in 1916. The concept was further developed as a process involved in self-organization by the theoretical physicist Hermann Haken (born 1927 in Leipzig), the biologist Peter Corning (born 1935 in Pasadena), and myself (born 1951 in Caracas). The figure presents data from “Google Books

Ngram Viewer" for the word “synergy”, ranging the years 1800 to 2000. This shows that the use of the word synergy is relatively recent, starting after the year 1900.

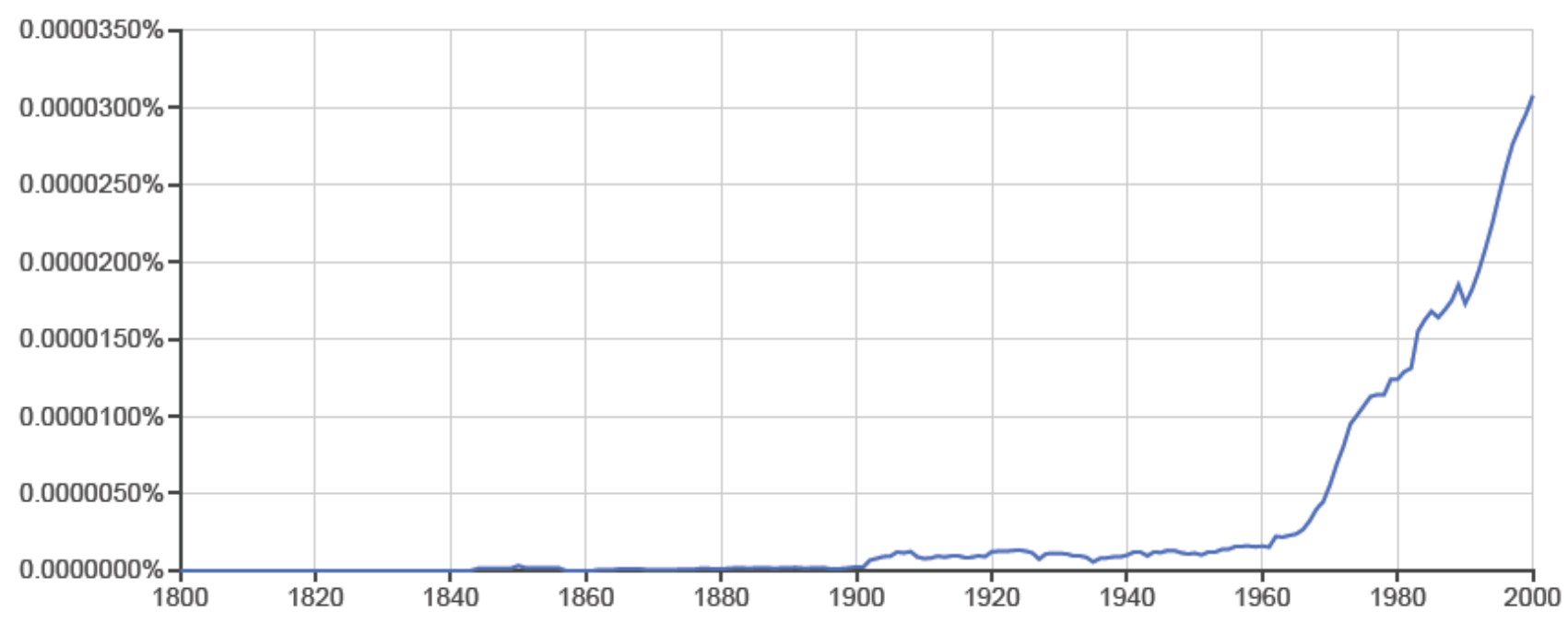

**Figure 2.1**

Frequency of the word “synergy” in the texts of Google Books for each year since 1800.

Source: Google Books Ngram Viewer

In the worlds of many components or constituent parts, complex interactions are not always arithmetically conserved. That is, two forces may add up to less than what the sum of each one would yield when measured separately because they dissipate energy when joined; they might conserve their energy perfectly and sum up arithmetically their energies when joined; or they might interact synergistically releasing more energy than what they both represent when measured separately.

This relationship is called non-linear in the sense that the relation between the variables, when plotted on a graph, do not follow a straight line. A linear relationship, for example, is that of an addition or a multiplication of two variables. With the appropriate transformation, they can be graphically represented in a two-dimensional plot with a straight line. This cannot be achieved with a non-linear relationship. The mathematical way of expressing that the outcome of an addition is different to the expected or is more than the sum of the parts.

Here we explore the nature of synergy in order to find its roots with the utilitarian intention of using this knowledge in our everyday life, in our business, in economics, in industry, in science, in politics, or wherever we are interested in improving our performance.

## 2.2 The Roots of Synergy

From the innumerable examples of synergy known to us in many areas of science, we can uncover the roots of synergy by focusing on the common elements in all those examples that are needed for the existence of synergy. Especially point 7 below may be considered essential in defining synergy, as it can be measured empirically. This hypothesis can be easily refuted by a counterexample (see Falsifiability by Karl Popper, born in Vienna 1902, died in Kenley, UK in 1994): find a case of synergy where some of these elements are absent, and the hypothesis has to be rejected. In the meantime, and after years failing to find such an example, we accept this hypothesis as true.

The fundamental elements or roots of synergy are:

1. Open stable systems
2. Cohesion
3. Diversity and Assortment
4. Complementarity
5. Communication and Synchrony
6. Freedom and Self-Organization
7. Dynamics of decreasing entropy increasing free energy

1. An open and stable system is required. It has to have some kind of membrane, border, porous boundary, or some force or motivation between the agents, that keep the parts together. Without a system it is impossible to speak of synergy. In closed systems, classical thermodynamics applies, and no increase in free energy with a simultaneous increase in negentropy is possible.

2. An a priori requirement in defining a system is that the system can be identified and measured. This requires that the system has boundaries, or as in the case of social systems, common characteristics among the members, or forces that aggregate and give cohesion to the parties.

3. There is often diversity or variety among the elements or characteristics that form the system; but not too much. Assortment or homophily (attraction to others with similar characteristics) seems to be important to reduce excessive diversity

4. The various component elements of the system must be complementary. Elements that repel each other cause the system to dissolve.

5. There must be some form of communication between the parties and with the environment to be able to synchronize

their actions and achieve complementarity in their performance. This communication may be direct or indirect and may have different forms. For example, science allows better communication with nature and social forces.

6. Freedom means here that order is the product of a dynamics that starts at the bottom and expands to the top of the complexity hierarchies. Top down instructions may also work synergistically as in the case of altruistic punishment. Every part of the system has some kind of function or effect on the whole (This is called “Ubuntu” in southern Africa)

7. The system is the result of a process where entropy decreases and free energy increases. The actions of the system can also be synergistic if they have these two thermodynamic characteristics. Entropy and free energy are the variables that are amenable to empirical and practical measurement, to evaluate the degree of synergy achieved or produced (see chapter 8).

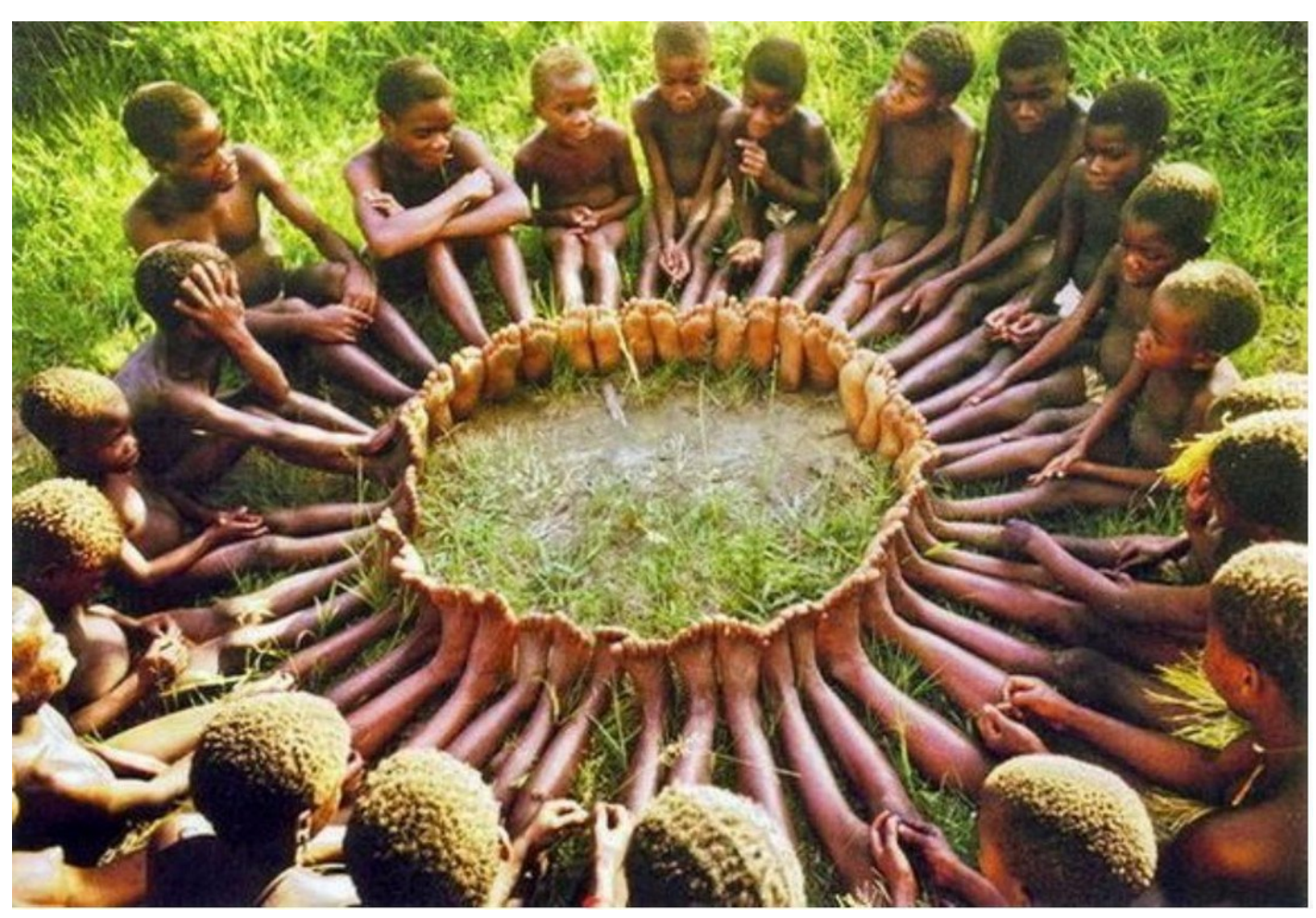

**Figure 2.** Ubuntu Ceremony

Creed in South Africa that a universal bond unites humanity. Source Google Images

## 2.3 Evolution and Synergy

Once a system discovers a synergistic arrangement, it is likely to remain. This triggers the evolution of synergy in systems from simple inefficient forms of organization to more complex and more efficient ones. Here I will give examples of these processes in economics and biology.

Synergy is not the only force of nature. Competition also drives evolution. The combination of competition and synergy achieved through cooperation seems to be the prime mover of most of the interesting processes that affect humans and their society. But between competition and synergistic cooperation, synergy is the least known of these phenomena.

## *2.4 Synergy in Economics*

Although evolutionary change in economic systems has been regarded as part of the dynamics of economics, traditional mainstream economics views the economic dynamics as an equilibrium between various forces, and thus envisions stable equilibrium states as the expected outcome of economic forces. Contemporary thermodynamics of complex systems, however, shows that equilibrium states in complex systems are not likely. Complex economies are thus far better described as dynamic evolving adaptive systems[2]. Chronologically speaking, the mathematical physicist Carl Neumann (born 1832 in Königsberg and died 1925 in Leipzig) is thought to have been the first to theorize about economics thermodynamically. Several other economists dwelt on thermodynamics[3]. Special notice is due to the economist Nicholas Georgescu-Roegen (born 1906 in Constanta, Romania and died 1994 in Nashville) who tried to build economic theory on irreversible (i.e. non-equilibrium) thermodynamics in his book The Entropy Law and the Economic Process.

Viewed through our lens of synergistic ideas, humanity has suffered several economic revolutions, each one thanks to novel types of organization that achieved synergies increasing human productivity to new levels of prosperity. One was the domestication of plants and animals leading to the adoption of agriculture that allowed the establishment of villages and cities.

This revolution was possible thanks to the synergies unleashed by the plant-animal-human symbiosis that allowed humanity overcome the limitations of nomadic hunting and gathering and exploit the benefits of sedentary life. Another series of economic revolutions were triggered by technological advances rooted on a novel conception of science[4]. The industrial revolution, the information revolution and revolutions triggered by robots and artificial intelligence, are all rooted on the synergies unleashed by novel technologies that could emerge thanks to our investments in science.

This view of the evolution of economics is not new, but neither is it main-stream. Understanding the details of the processes of how synergies emerge between human labor, human needs, markets and technology will be fundamental in maintaining prosperous economies that sustain healthy societies in the future. It is time to take such studies more seriously.

Synergy might also be negative in economics and elsewhere. A good example is the financial pyramid scheme that synergizes human folly by exploiting the addiction of our minds to simple and frugal thinking. Here, the system grows non-linearly with more and more participants paying into a fund, until it suddenly collapses by not being able to pay newcomers. Simple and frugal thinking is often efficient but has its shortcoming driving addiction to games, charlatanism and dogmatism. Synergy between dogmatism, resentment and ignorance can lead to negative economic growth.

The phenomena that produce wealth can be dissociated from those that produce poverty. The roots of poverty are not the lack of wealth. Poverty has its origins on its own negative synergistic processes or in an absence of synergies that produce free energy at the level of the individual and society. It is this useful energy that allows the production of goods and services, that underpin the wealth of individuals and of society. Poverty is rooted in attitudes and circumstances that block the formation of synergistic relationships, thus preventing the accumulation of wealth.

The drive to accumulate goods and the hoarding of resources are instincts that we share with many animals. These instincts promotes the capitalist business spirit but also theft, corruption and war. It is the synergistic management of these instincts, along with others such as love, shame and responsibility, which lead to economically successful cultures or alternatively to societies incapable of surpassing primitive models of coexistence and production. By knowing and managing the various factors that are at stake, we will be able to unleash potential synergies and advance towards increasingly productive and inclusive economies.

## 2.5 Synergy in Biology

Peter Corning writes in 2005[5] "*it is the "payoffs" associated with various synergistic effects in a given context that constitute the underlying cause of cooperative relationships -- and complex organization -- in nature. The synergy produced by the "whole" provides the functional benefits that may differentially favor the survival and reproduction of the "parts". Although it may seem like backwards logic, the thesis is that functional synergy is the underlying cause of cooperation, and organization, in living systems, not the other way around. So it is really, at heart, a functional and "economic" theory of emergent complexity in evolution, and it applies both to biological and cultural/political evolution.*"

Peter A. Corning and Eōrs Szathmáry (born 1959 in Budapest) in their article "Synergistic selection": A Darwinian frame for the evolution of complexity [6], apply this definition of synergy and propose that synergistic effects of various kinds have played an important causal role in the evolution of complexity, especially in the "major transitions". The synergy achieved in each transition achieved otherwise unattainable functional advantages arising from various cooperative phenomena. These "Major Transitions in Evolution" have been described by John Maynard Smith (born 1920 in London and died 2004 in Lewes, UK) and Eörs Szathmáry [7] and include:

1. The emergence of the first replicating molecules in segregated, protective "enclosures" we now recognize as cells.

2. The origin of chromosomes, which linked various replicating molecules together in cooperative relationships in what is now called a genome.

3. The origin of the genetic code for protein synthesis that linked RNA-based auto-catalysis involving DNA

and proteins.

4. The origin of eukaryotes from independent, free-living prokaryotes.

5. The rise of sexual reproduction.

6. The emergence of multicellular organisms.

7. The origin of social groups culminating in complex, highly integrated, communications-dependent species with a social division of labor, like honey bees and humans.

Each of these transitions involves the cooperation of different partners, each specialized in a different means to harvest nature. This cooperation achieves synergies that make them hugely successful, shifting the odds of evolutionary adaptation to favor their selection. In this view, irreversible synergistic processes define the evolutionary path of all living organisms and structures.

Case 5 or the emergence of sexual reproduction might make the point. Sex is ubiquitous among living organisms, especially among multi-cellular organisms such as plants and animals. The emergence of sex is considered one of the major transitions in evolution, but theoretical biology struggled with this issue for a long time. Our understanding of the evolution of sexual recombination is still very partial and incomplete. Several important concepts have been used to explain the ubiquity of sex and its success in biological evolution, but none passed the experimental tests of computer simulations. A popular proposition suggests that sex accelerates evolution to maintain species one step ahead of parasites, the so-called Red Queen hypothesis[8], or that sex uncouples beneficial from deleterious mutations, allowing selection to proceed more effectively with sex than without it[9], does not explain the emergence of sex in computer simulations. Computer experiments[10] showed that the evolution of sex can be much better explained by just assuming the existence of synergy between the sexes. Looking at this synergy opens a treasury of examples and diversity of cooperative strategies. Biologists and sociologists have focused more on the conflict between the sexes forgetting the basis upon which they work: synergistic cooperation.

## 3. Science correcting our Ego

To dig more efficiently into synergistics, we must understand the relation between knowledge and synergy. This requires us to recall some basic scientific knowledge. I will try to do it as painlessly and intuitively as possible.

The **ego, me** or **I**, is a diffuse entity, which includes our body, our mind, and our future transcendence. It can be viewed as an algorithm - or set of rules - of our working mind, selected through biological evolution, that is fundamental for our survival. Our ego, however, sometimes takes us into blind alleys. Thomas Hobbes (born 1588 Westport Wiltshire; died 1679 Derbyshire UK) wrote about "*a vain conceit of one's own wisdom, which almost all men think they have in a greater degree than the vulgar; that is, than all men but themselves, and a few others, whom by fame, or for concurring with themselves, they approve.*" and he adds that it is only a "*special skill of proceeding upon general, and infallible rules, called science; which very few have*" that allows humans to overcome this shortcoming. Today we know that the basis of the effectiveness of science is the recognition that our mind is very limited in grasping complex interactions and that empirical evidence is the last arbiter in our quest for knowledge. Modern science recognizes that empirical experimentation or observation is the most efficient device to overcome the limitations of our mind. In this view, irrationality is the rejection of empirical evidence as a tool for advancing knowledge. The aim of science is not to find the absolute truth, as such a thing probably does not exist, but to increase and improve our knowledge of ourselves and of the surrounding world. This surrounding world includes complex phenomena and processes where synergistic interactions emerge.

Here I will try to keep as close as possible to science as to human common sense and intuition. When this is not possible, I will give priority to science. The reader who endures this difficulty will be rewarded with a deeper understanding of synergy and other related processes. However, the rationale behind synergy can be grasped without being a scientist. There are many ways to understand synergy. If one explanation is not grasped easily, a different one might. The readers should skip sections they find intractable. Other sections might be more amenable to their personal experience.

Rational thinking has evolved by adaptation. This has led our mind to adapt ways of thinking that are fast and frugal. Fast to react efficiently to urgent dangers, and frugal to spend less time and energy in producing them. Complex problems often are nor tractable with fast and frugal thought. They require more detailed analysis and deeper explorations. That is when science is needed most. Therefore, intuition alone will not help us in understanding complex problems.

### *3.1 Complexity and Emergence*

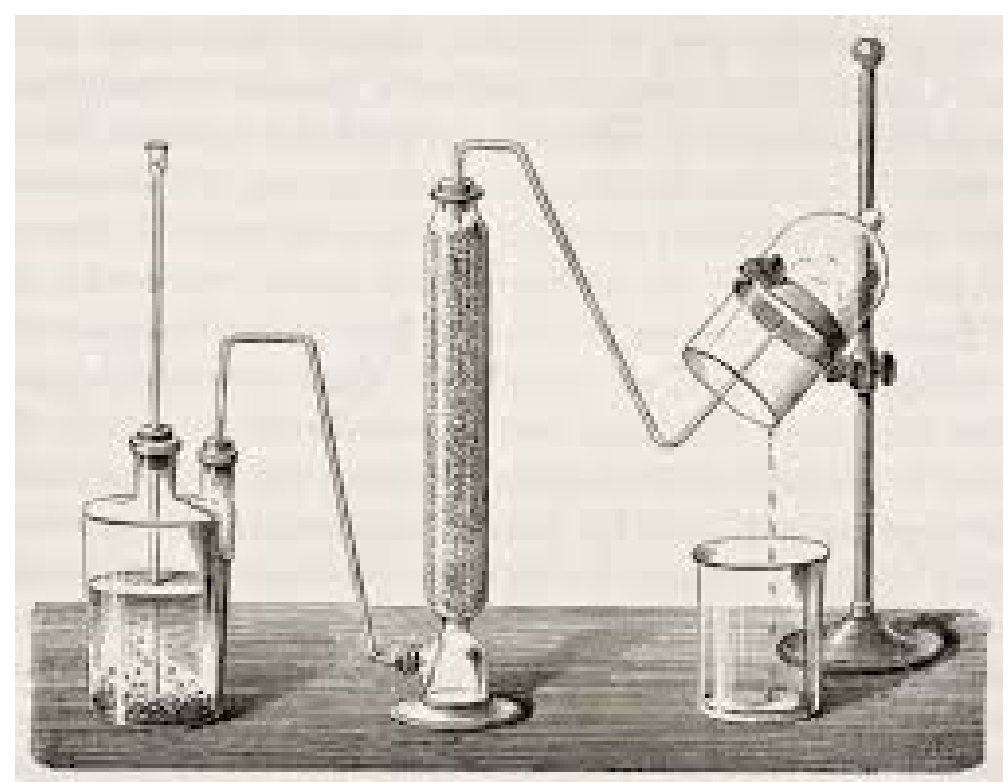

**Figure 3.1**

Ancient equipment to achieve the chemical synthesis of water molecules, starting from atoms of hydrogen- and oxygen-gas. Source: Google Images.

Synergy is a phenomenon that is only visible in complex systems. That is, mathematicians will never claim that 1 + 1 = 3 (see exceptions in [11] ). But they also will never attempt to add oranges and pineapples, or hydrogen and oxygen. Yet for chemists who deal with more complex abstractions, adding different elements or compounds together is a routine occupation. Chemists seal hydrogen gas and oxygen gas in a container and pass high fluxes of energy through the mixture producing liquid water. Adding up elements in chemistry inducing chemical reactions produces new compounds that have completely different properties to the isolated constituting elements. These processes have no simple mathematical description but can be analyzed and described with tools that we now call chemistry, thermodynamics and quantum mechanics.

Other reactions involve the formation of hydrogen molecules from individual atoms that start sharing electrons represented in the equation:

$$H + H = H_2$$

Or in a nuclear reaction the fusion of two hydrogen atoms to produce helium described in the equation:

$$H + H = He$$

Clearly, abstract mathematical representations describe different phenomena where the meaning of the symbols “+” and “=” differ.

In the first equation, + indicates a simple addition of identical objects or entities and synergies cannot appear.

In the second example + symbolizes the addition of different elements or substances in a chemical reaction. In this case, the component part of the addition interacts physically and chemically in a chemical reaction, rearranging themselves – in this case their electrons – to form a new substance.

In the third equation, we represent a nuclear reaction. Here, with an enormous release of energy, the nuclei of the two Hydrogen atoms fuse to form a new element: Helium.

### *3.2 Energy and Entropy*

In colloquial terms we can say that energy is an action force or physical entity that has the capacity to do work. That is, the concept of energy is defined as the ability to produce work. In any process, this energy can be subdivided into two parts: the one that actually performs the work or Free Energy (***Φ***) and another that is dissipated and lost for use in future work, which we call Entropy (***ἐ***). The opposite of entropy, or negative entropy (***-ἐ***) we call Negentropy (***ŋ***). This negentropy is the opposite of dissipation and disorder and represents order and information.

During the beginning of the scientific revolution, the engineer Sadi Carnot (born 1796 in Paris and died 1832 in Paris) worked out a formula for how efficiently steam engines can convert heat into work to push a piston or turn a wheel. Heat is now known to be a random, diffuse kind of energy and there are many ways to convert it to an orderly kind of energy. To Carnot's surprise, he discovered that a perfect engine's efficiency depends only on the difference in temperature between the engine's heat source (typically a fire) and its heat sink (typically the outside air). Work is a by-product, Carnot realized, of heat naturally passing to a colder body from a warmer one. His efficiency formula evolved over the 19th century into the theory of thermodynamics: a set of universal laws dictating the interplay among temperature, heat, work, energy and entropy — a measure of energy's incessant spreading from more to less energetic bodies. The laws of thermodynamics apply not only to steam engines but also to everything else: the sun, living beings and the entire universe. This idea that energy has two forms, one which is useless heat and one called free energy that produces useful work, is a cornerstone of modern thermodynamics. However, the type of energy under study might be very different. Thermodynamics applies to all of them, from the energy attracting nuclear component, gravity, electricity, magnetism; the more derivative types of energy such as chemical, biochemical; up to financial types such as monetary resources.

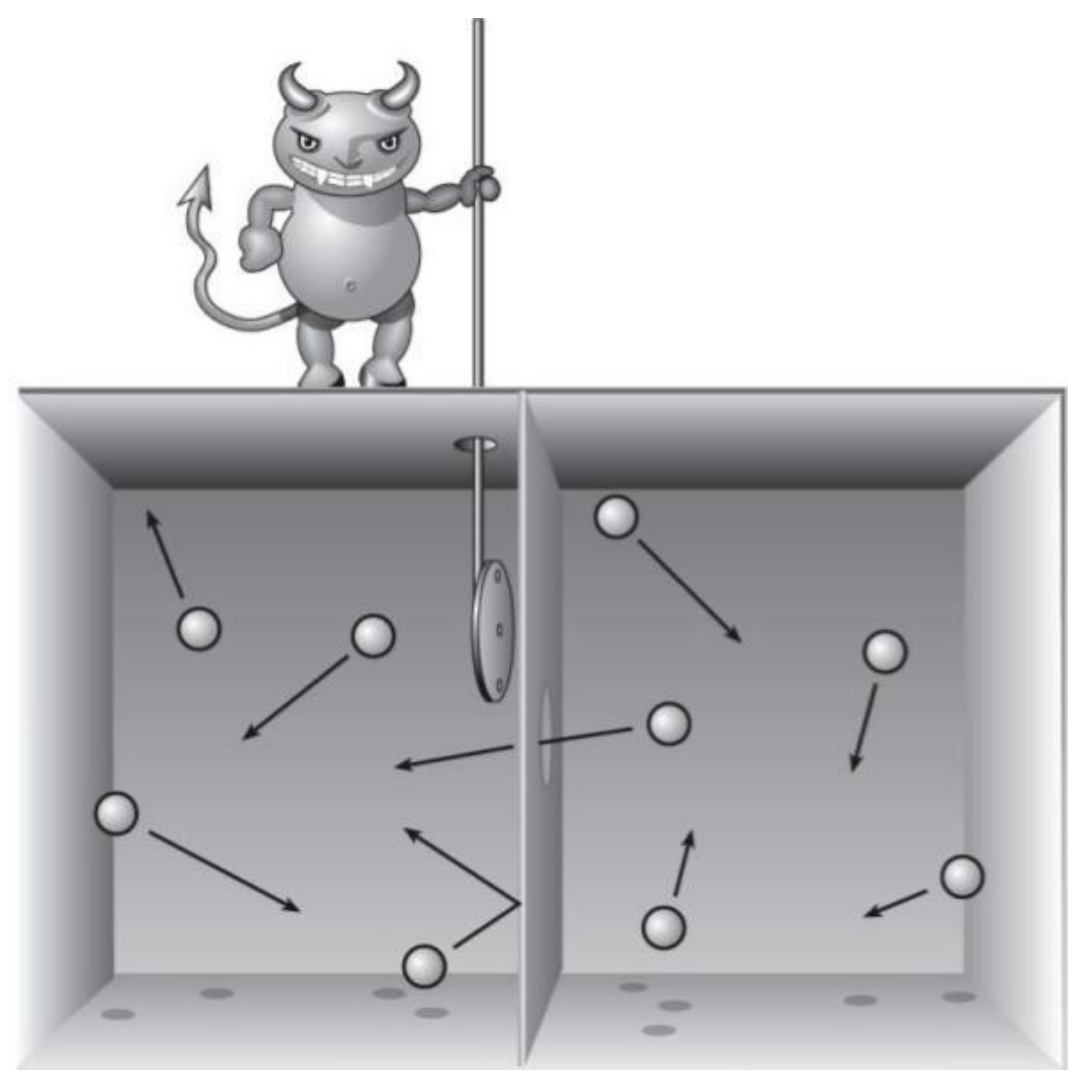

**Figure 3.2a**

A demon separates fast moving gas molecules from the slow-moving ones, storing them in different compartments. (Maxwell's Demon: Scientific American Blog Network )

Another important modern thermodynamic insight is the nature of information. The mathematician James Clerk Maxwell (born 1831 in Edinburgh; died 1879 in Cambridge-UK.) described a thought experiment in which an enlightened being — later called Maxwell's demon — uses its knowledge to lower entropy and violate the second law by producing useful work from normally non-useful dissipated energy. In a container of gas like a soup of randomly moving particles, each with different speed or kinetic energy, the demon knows the positions and velocities of every molecule. By partitioning the container and opening and closing a small door between the two chambers, the demon lets only fast-moving molecules enter one side, while allowing only slow molecules to go the other way. The demon's actions divide the gas into hot and cold, concentrating its energy and lowering its overall entropy. The Maxwell Demon reverts the natural tendency for heat to flow from hot to cold parts of a body or gas soup. The once useless gas can now be put to work.

Continuing this reasoning, the physicist Charles Bennett (born 1943 in New York City), building on work by Leo Szilard (born 1898 in Budapest; died 1964 in La Jolla) and Rolf Landauer (born1927 in Stuttgart and died 1999 in Briarcliff, NY), linked thermodynamics to the young science of information. Bennett argued that the demon's knowledge is stored in its memory, and memory has to be cleaned, which takes work. The overall entropy of the gas-demon system increases, satisfying the second law of thermodynamics, and showing that information and energy are related concepts.

Energy, in its multiple forms, is an element that is present in all complex interactions, and that is not represented symbolically in the equations discussed above. To understand the working of these complex phenomena we need to consider the energy consumed and the energy released by these processes.

In doing so, we need to rewrite the equations as follows:

$$2\ H + O + \Delta = H_2O$$

and

$$H + H + \Delta = He + \Delta$$

where Δ represents heat or other forms of energy.

In the case of the nuclear fusion of Hydrogen atoms, we need energy to induce the fusion, and the fusion reaction produces energy. If the energy produced is much larger than the one consumed to trigger the reaction we might conclude that this resembles what we had defined as synergy. But in the case of fusion of Hydrogen atoms, the energy released was formerly stored as potential energy in the nuclear structure of the Hydrogen atoms; so that the total energy (potential + free energy) before and after the reaction is the same. This is an example of the fact that synergy cannot be produced in a closed system. The laws of thermodynamics apply to closed systems. That is, in a closed system energy cannot be created nor does it disappear, it is transformed to other kinds of energy or to heat or entropy.

The best illustration of the difference between closed and open systems is life. Living organisms must consume food, water, and air to grow. As soon as the organism isolates itself from the environment, becoming a closed system, it stops consuming the vital elements and dies, reaching thermodynamic equilibrium.

There are several different kinds of energy. One, defined in the 1870s, by mathematician Josiah Willard Gibbs (born 1839 in New Haven; died 1903 in New Haven) and physicist Herman von Helmholtz (born 1821 in Potsdam; died 1894 in Charlottenburg), is called “Free Energy”. This is the available energy that can be used to perform work. This energy can often be stored in the nuclei of atoms, or in the electrons that form chemical bonds holding together chemical compounds and molecules, or as water stored in a dam high up in the mountains, or in a diversity of other ways. This stored energy, we call potential energy, and can eventually be recovered to perform useful work. Another kind of energy is heat that is diffusely distributed in a body or a system. Another kind of energy exists that cannot be used any more to produce work. This energy feels more like noise or random disorder and is called Entropy.

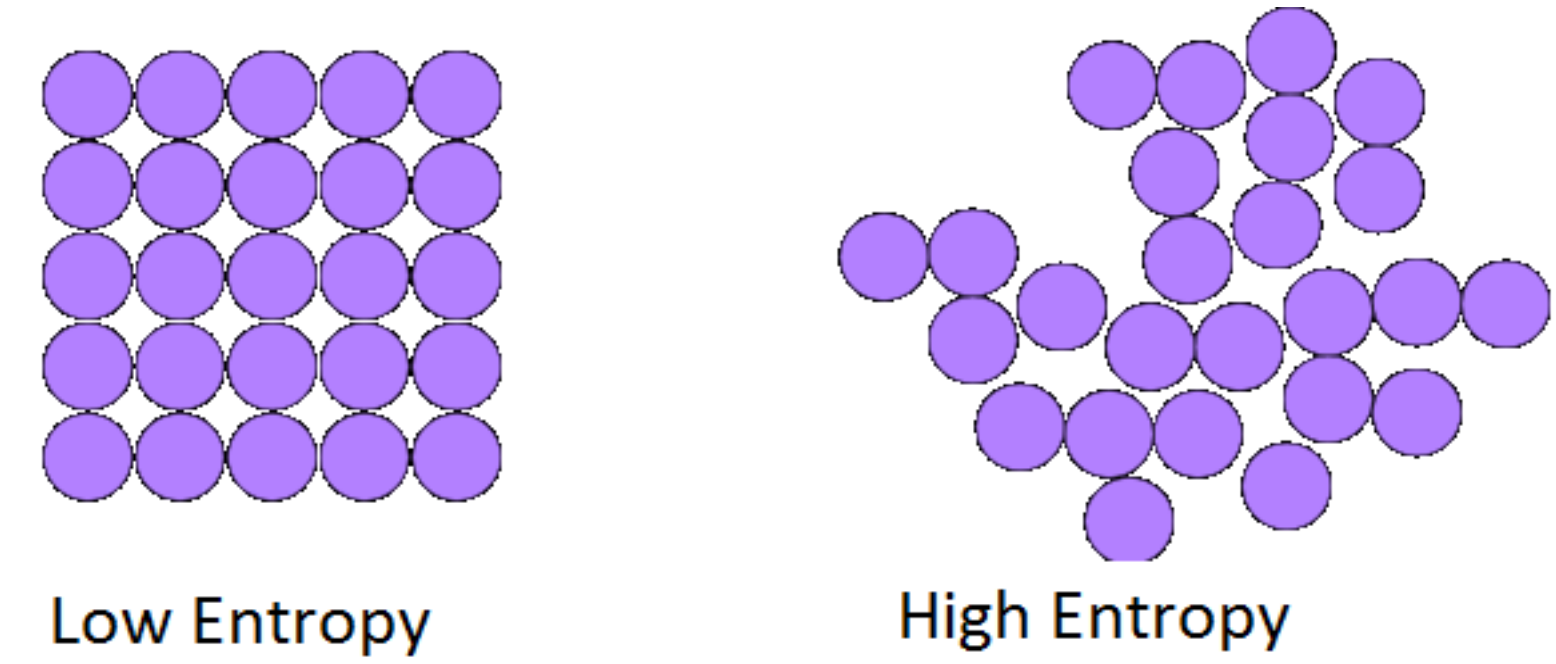

**Figure 3.2b**

Two systems with different entropy levels .

(drawing by the author)

Low entropy represents high negentropy, and high entropy represents low negentropy.

Nature not only provides us with stored energy, but also with fluxes of energy. A waterfall dissipates energy continuously just as water molecules, attracted by earth's gravitational field, drop from the cliff. Fires emit energy in diverse forms, as the energy of chemical bonds of the burning compounds is released. Living organisms burn stored metabolic energy to maintain themselves and allow for a diversity of actions. These complex processes occur in a so-called "open system" where systems are not completely isolated from their surroundings and the interchange of energy and information is possible. In these systems, the second law of thermodynamics does not apply, as the system can absorb energy, matter or information from the environment and discharge heat and entropy to the outside of the system, avoiding the ever-increasing accumulation of entropy inside, as demanded by the second law of thermodynamics for a completely closed system.

To grasp these processes, we need to rationalize our intuitive understanding of the relationship between open systems and complexity. This will be far easier than the intimidating words and concepts presented. We will open a wonderful new world.

## 3.3 Systems far from Equilibrium

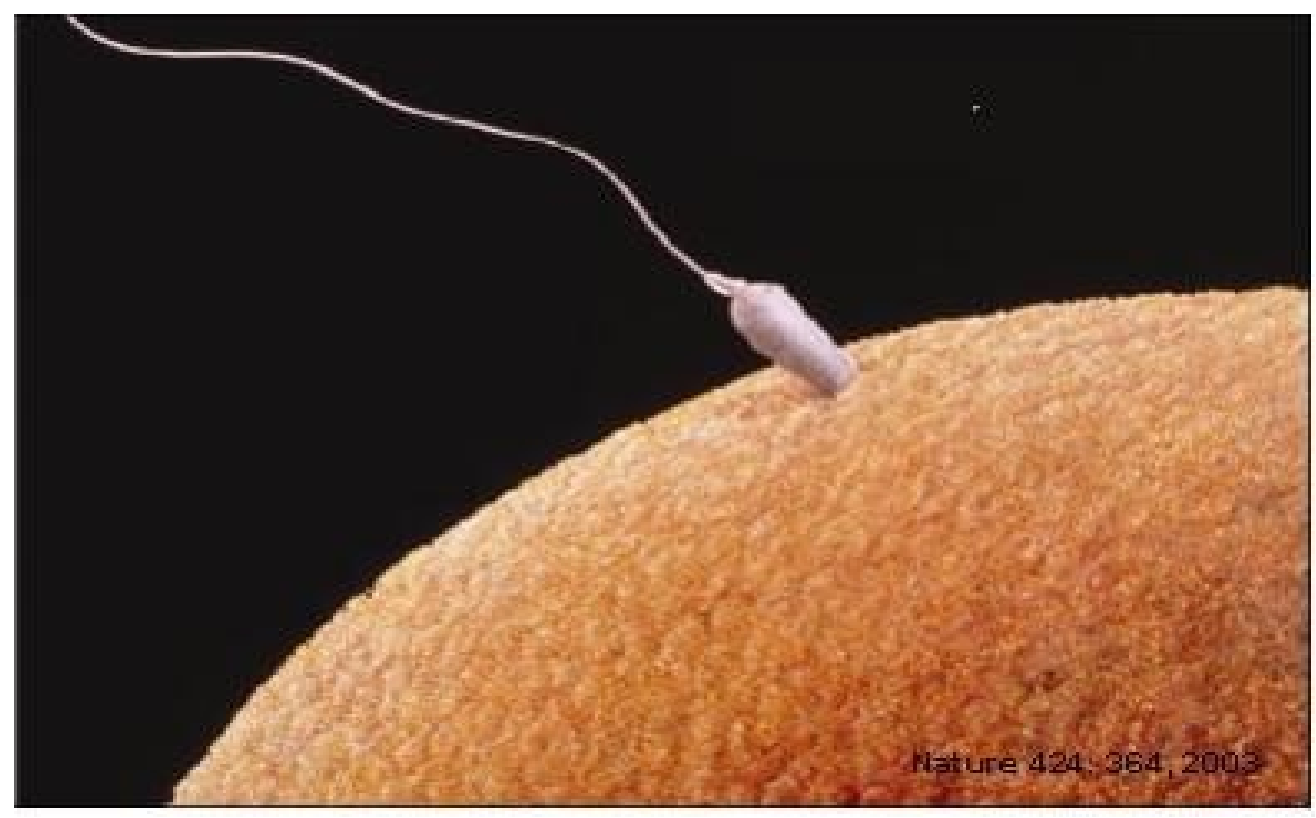

**Figure 3.3**

Fertilization in action: A sperm is penetrating an egg.

Source: Journal Nature

Most systems, dear or important to us, are thermodynamic “open” systems, far from chemical or physical equilibrium. That is, the system is not completely isolated from its surroundings, and fluxes of matter and/or energy between the system and the surroundings might occur. This allows for processes to show nonlinear or non-additive properties. For example, a cell in our body, a living organism in an ecosystem or a human society surrounded by other societies, are all systems that suffer continuous change (they are far from equilibrium) and have important exchanges of energy, nutrients and other elements with the surrounding environment (they are open). In the figure, we show a human sperm in the process of penetrating a female ovum during fertilization. Analytical mathematics is quite useless in describing this process. Writing S + O = N symbolizing that a sperm added to an ovum produces a new living being, is of no great help in understanding what is happening here. Open systems far from thermodynamic equilibrium require new tools for its analysis. These are being produced by a field of science called irreversible thermodynamics.

An important feature of these systems is that they engage in irreversible processes. That is, once an egg breaks, it is not possible to reassemble it. Thus, equations using “= “can't represent irreversible processes. These processes are ubiquitous. Water spilling from a beaker, toothpaste squeezed out its container, living organisms dying, or empires collapsing, are all examples of irreversible processes; but so are the growth of plants and animals, the accumulation of wealth, and the process of learning.

In physics, we associate an increase in information complexity with a decrease in entropy. The negative or opposite of entropy is negative entropy and is called "negentropy". In the example of the production of Helium, negentropy increases (entropy decreases) as Helium atoms can be considered to be more complexly structured than Hydrogen atoms. Erwin Schrödinger (Born 1887 in Vienna and died 1961 in Vienna), the 1933 Nobel Prize winner in Physics, in his book "What is Life", published by the Dublin Institute for Advanced Studies at Trinity College, proposes that life is a system that needs to eat some kind of order (negentropy) and energy to maintain itself alive. That is, life is an irreversible process that occurs in open complex systems. In such systems, as recognized by Ilia Prigogine (born 1917 in Moscow and died 2003 in Brussels), who won the 1977 Nobel Prize in Chemistry, no equilibrium states exist. But many states are like those at equilibrium which he called Stable States. These stable states, such as a living cell inside an organism, show some predictable properties that change little in time, such as the flux of nutrients from the environment to the cell, and a flux of metabolites or decomposed products from the cell to the outside.

In terms of the physics and thermodynamics of stable states in open systems we now know that:

1. Many complex systems draw their energy to maintain a given stable state from a flux of energy or from a pool of potential energy at an external source. Many different types of energy and ways of harnessing energy exist. For example, the source of energy in hydroelectric power stations is the flux of water due to the pull of gravity. The source of energy of ticks is the blood of the host. And the rose-bush draws its energy directly from the sun, thanks to the photosynthetic capabilities of its leaves.

2. Synergy occurs when two entities collaborate in harnessing a flow or pool of energy or of information, collecting together more energy than what neither of them could have achieved on its own. Proteins and several other molecules transform the energy of light to chemical energy in the photosynthetic process in the rosebush leaf. Several different mechanical components allow the movement of water to be transformed into electricity in the dynamos of a hydroelectric power station. Several anatomic features, such as penetrating styluses and pumping devices in the tick's body, together with several specific behaviors, allow ticks to suck blood from their hosts.

3. Synergistic processes not only increase the available free energy of the system, they also increase the negentropy (decrease the entropy) of the system. More efficient tick species, power stations and rose species have more sophisticated arrangements for their specific functions. For example, ticks, and blood sucking mosquitoes have a set of receptors to detect their blood carrying hosts that no other organism possesses.

4. In social settings, the more varied and diverse the individuals, the greater the range of tools for harnessing energy, the less interference between them, the greater the synergy they can achieve. Among nations, the countries with the highest electricity consumption per inhabitant have also the highest production of academic articles; the universities with the highest diversity of academic formation, nationalities and race among their staff and students; the more diverse supply of consumables and the more sophisticated institutions to run private and public interests.

It is in irreversible processes involving open systems in or near a steady state, such as those characterizing living organisms, ecosystems and society, that synergy is most likely to occur. A series of examples will be presented to grasp intuitively how synergistic processes work, but before that, we need to address another fundamental dimension of reality: Time.

## 3.4 Time Horizons

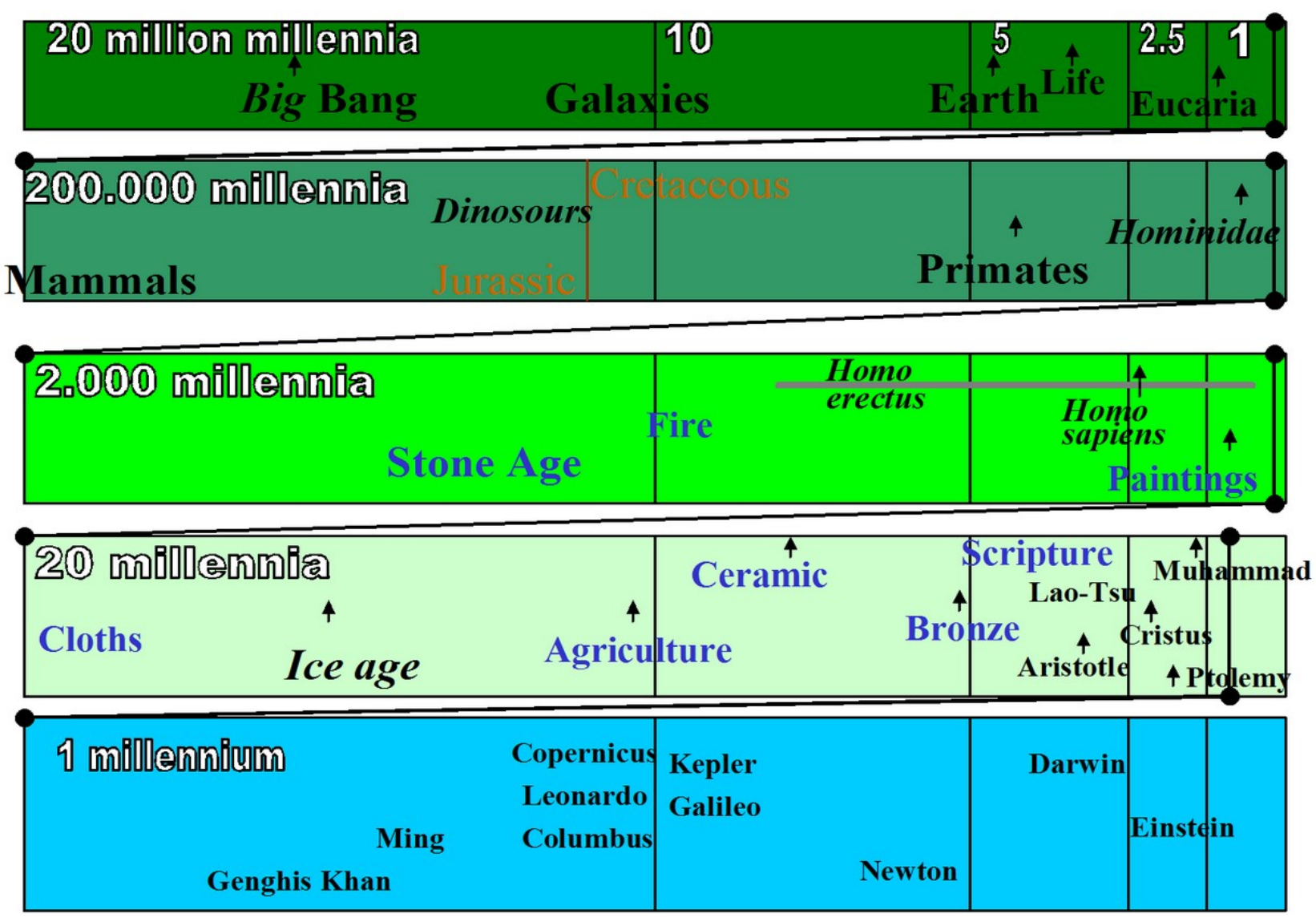

**Figure 3.4**

Our time windows to watch the world. Source: drawing by the author

Depending on which time frame we choose to view history, we will discover different laws that rule the temporal dynamics in that time window. Some examples are given in Figure 3.4. Here we observe the emergence of order in the 20 million millennia window, biological evolution in the 200000 millennia window, and the ascent of humanity (i.e. of *Homo sapiens*) in the 2000 millennia window. Each of these time windows are studied by different scientific disciplines. Descending from the upper top window down to the fifth at the bottom in the figure, we can list Cosmology, Evolutionary Biology, Anthropology, Archaeology and History as the disciplines most relevant for each of the time windows. The interesting fact is that in each of those time windows, synergistic processes occur. Remarkable is the formation of stars and emergence of galaxies in the 20 million millennia windows; the emergence of primates in the biological evolution of mammals in the 200.000 millennia windows; the emergence of the modern human species in the 2.000 millennia windows; the emergence of sophisticated knowledge societies and cultures in the 20 millennia windows; and the emergence of modern science during the last millennium.

Certain actions and aims may seem rational in each time horizon but are irrational if viewed in a different time horizon. Thus, it might seem rational to eat as much food as we can in a single day if we do not know if we will have access to food the next day, but it is completely irrational to eat as much as we can every time we have access to food if we know that food is plenty and that in the long term, we will become obese and die early. This allows certain processes to have long term synergistic effects that are not visible in shorter time windows. So, for example, an altruistic act if looked at in a short time horizon, might seem to be a sacrifice by an individual to benefit another, but viewed with a large time horizon, that same act might be better described as a social investment that benefits both actors eventually.

But even if looking through the same time window, different perspectives of the same phenomena might be glimpsed. Synergy is often related to emergence, self-organization, stigmergy, synchrony and other complex phenomena. Some of them have nothing to do with synergy, such as chaos and fractals, though they are very interesting on their own.

# 4. Nature's way to achieve Synergy

Synergy is present in many domains, ranging from physical processes, chemical reactions, biological phenomena, engineering, and social interactions among humans and among animals. Let's analyze some examples in detail to get a better understanding of synergy. Some historical important examples are complemented with descriptions of contemporary social processes that depend on a better understanding of synergy for their future improvement.

## 4.1 Adaptive value of Liberty

Definitions of Liberty:

- "The power or right to act, speak, or think as one wants without hindrance or restraint" (Oxford)
- *"Freedom is the ability to act according to one's own will"* (Wiki)

None of these definitions are independent of our subjective feelings. Neither is the "free will" promoted by Arthur Schopenhauer (born Gdansk 1788, died Frankfurt 1860). These definitions are incompatible with what we have learned from neuroethological studies. Our “will” is conditioned by our genes, adapted to optimize the responses to our needs in the last 4+ billion years. Our motivations and “will” are formed and subjected to the action of genes, environment and culture. It is an epigenetic phenomenon. In this ethological context, freedom refers to the independence of the individual's actions that produce variations and allows evolution to occur. Variety is the substrate of natural selection. Individual freedom allows variety to exist.

No absolute objective definition of liberty or of freedom exists. This causes important confusions and misunderstandings. Example of conflicts regarding the relative exercise of freedoms are:

Conflict 1: Freedom in partner selection or sexual conflicts. In this situations, different actors will optimize different kinds of freedom. The freedom of the future mother in choosing the partner of her children optimizes status and social structure of the mate to secure resources. The future father might be interested in the child's long term health regarding the biological workings of the immune system. Biologically, love optimizes genetic "assortation" for a fitter offspring.

Human solutions to these dilemmas include age limit for marriages, sex education and multiple cultural norms

Conflict 2: Freedom to use natural resources promotes monopolies, reducing the efficiency in the sustainable use of resources. At the same time, it promotes investment, increasing the efficiency in the use of resources. Human solutions to these conflicting freedoms are Social Democracies, Sustainable Capitalism, Liberal Ethics and other cultural contraptions.

Conflict 3: There is no freedom without law and order, nor law and order that does not limit freedom. Humans solve this problem by authoritarianism and arbitrary imposition of laws with no long term stable equilibrium in sight

A different perspective is provided by physics with the concept of “degrees of freedom”. This allows freedom to be quantified in physics, statistics, classical mechanics, thermodynamics and other areas of science. The degree of freedom refers to the number of independent coordinates (dimensions) necessary to determine the position of an entity. This requires accepting the multidimensional nature of reality (manifold order). Reality is not as simple as our mind, using fast and frugal algorithms, forces us to see it. By operationalizing the concept of freedom to a specific situation, multiple ways of seeing it appear. In addition to physical dimensions we know of degrees of freedom in the Bio-Socio-Economics space. Some examples are:

- In the stone age we exercised more freedom of action in many aspects than in modern society, but natural forces limited us more. The dimensions in which we can act today are more numerous, opening novel and unknown possibilities of actions, despite the loss of some freedoms of the past. We do not normally walk around naked anymore nor do we kill anyone we do not like.

- To travel the world we have to temporarily limit our freedom of movement on an airplane. We are confined in proportionally less space in an airplane than what is legal for chicken in an egg factory.

- To participate in a congress we have to shut up and listen to others

- To enjoy living in a clean and organized community we have to comply with rules and not spit nor defecate freely.

- To explore cyberspace we have to learn new languages that have new dimensions to knowledge.

- New forms of property limit our freedom as in the case of Patents, Radioelectric spaces, Records, Rules and contracts with or without consent, Monopolies.

- Pandemics and contagious illnesses limit our biological and social liberty. Refusing vaccinations against a pandemic agent limits the free life of others.

All these dimensions of liberty are in the realm of what we intuitively understand as negative entropy. In this respect, evolution is an irreversible process. Historical or evolutionary decisions reduce freedom (increase negentropy) but increase free energy. This is what I propose as a fourth law of thermodynamics (see section 4.2). That is, gains in Free Energy or Useful Work can only be achieved by reducing Entropy. The reduction in entropy implies an increase in negentropy or of a certain type of order or information in different dimensions of the knowledge space.

More advanced and sophisticated societies require more specialized individuals, with less freedom of economic action. This has been shown to be true in ants, rodents, humans, and other mammals. For example, in a hunter-gatherer society, each individual acts independently of what others do. In an agricultural society, individuals restrict their movements, become sedentary, and cooperate with others to produce food.

In a modern society, the synergies of social interaction are the basis of production and force an increasingly fine division of labor, reducing the freedom of action of the individual but increasing their potential for life and enjoyment of it. This is the product of the evolutionary process that increases degrees of freedom in new dimensions such as:

Better health and longevity

Greater movement capacity

Greater ability to access information

More diversity in consumer products

More and better tools to explore nature

Better handling of emotions

More sense of security

More intense interpersonal relationships

Thus freedom implies the ability to make changes, as explored by Bejan in the Constructal Law (see below), but changes limit some of our freedoms.

This way to understand freedom has consequences in the Social Economy as it forces us to recognize the Importance of Social Justice in Liberal Economics. The traditional Liberal focus assigning the right to property as the sole guarantor of freedom is not sustainable in absolute terms. Liberalism of the 21st century explores possible successful strategies in maintaining freedom in the various realms of human interests while advancing our capacity to produce useful work. These are specially relevant in Information Technology, Globalization, and novel democracy that liberate the individual from uncontrollable influences. For example, a fundamental flaw of humanity is that “*fools and fanatics are always so certain of themselves, and wiser people so full of doubts*”. Freedom of actions, in a democracy compound this problem. Possible strategies in promoting freedom while advancing our capacity to produce useful work include:

- Systems that empower the citizen (kleros.io)
- Economic freedom from Central Bank (cryptocurrencies)
- Moral freedom (gender diversity)
- Freedom of ideas (open networks and free preprints)
- Freedom of action (tolerance for new activities)
- Freedom from manipulation (Demarchy or lottocracy)
- Rational Democracy (Decentralized state with maximal self organization)

### *4.2 Free Energy and Negentropy*

A fourth law of thermodynamics [12]

The relationship between Free Energy and Negentropy can be visualized by a cannon ball placed upon a heap of fire powder. The ball will hardly move a few meters when the powder is burned. But if the fire powder is placed into a cannon and the cannon ball placed on top of the powder, the distance traveled and the speed of the flying cannonball will be very large indeed.

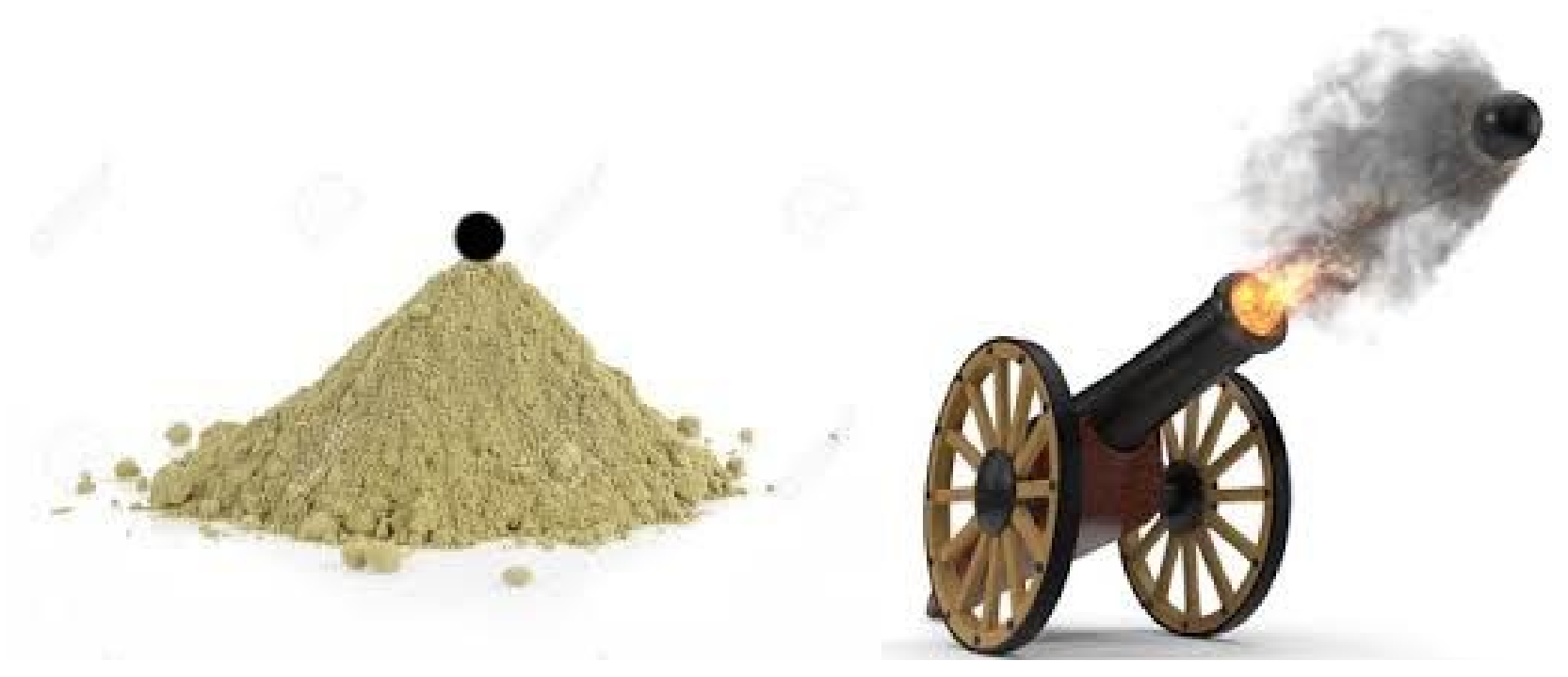

**Figure 4.2:**

Fire powder and cannon

This descriptive exercise lets us understand the difference in Free Energy ***Φ*** (spatial reach and kinetic power) of the cannonball in both situations. The difference of the border conditions or “order” regulating the explosion and propulsion of the ball in both situations is also enormous. One scalar of the equation is Free Energy ***Φ*** quantifies the amount of useful work that can be performed; the other scalar is the order, information or border conditions that are implicit in the design of the canon which we will call Negentropy ***ŋ*** . Therefore, a thermodynamic law regulating both scalars would state that increases in Free Energy Φ can only be achieved by increases in Negentropy ***ŋ*** . This law is not symmetric, not all increases in negentropy lead to increases in free energy [13]

This same logical relationship can be applied to the thermodynamics of social systems, and that of any complex system that is open to fluxes of mass and energy from the environment and is not in permanent equilibrium. Thus, we can propose a law of irreversible thermodynamics.

In other words, synergy emerges from synchronized reciprocal positive feedback loops between a network of diverse actors. For this process to proceed, compatible information from different sources synchronically coordinate the actions of the intervening agents resulting in a nonlinear increase in the useful work the system can manage. In contrast noise is produced when incompatible information is mixed.

In thermodynamic speak, this synergy is produced from gains in free energy density $\boldsymbol{\Phi}$ and of information density (negentropy $\boldsymbol{\eta}$ ). The gains in $\Phi$ reflect an increase in individual autonomy of the agents, with increased emancipation from the environment, with increases in productivity, efficiency, capacity for flexibility, self-regulation and self-control of behavior through a synchronized division of ever more specialized labor, and increased potential of producing useful work. Several examples are known of complex systems that provide quantitative data reflecting this fourth law of thermodynamics. This law of irreversible thermodynamics, states that for synergy to occur, increases in $\Phi$ require increases in $\boldsymbol{\eta}$. That is $\Delta\boldsymbol{\Phi} = k_i \cdot \Delta\boldsymbol{\eta}$ ; where k is a constant dependent on conditions i. This law only applies to open systems, as in closed systems the first law of equilibrium thermodynamics states that $\boldsymbol{\Phi} - \boldsymbol{\eta} = k$. Quantitatively, $\boldsymbol{\Phi}$ can be measured as work, free energy or potential energy; whereas $\boldsymbol{\eta}$ can be measured as entropy and/or as information.

## Entropy and information

The relationship between negentropy and information is not smooth. Both concepts relate to similar aspects of nature, but they are not identical. Depending on the language used to transmit the information, sometimes negentropy is not proportional to information. Thus, definitions, context and concepts used to describe negentropy need to be specified before making extrapolations to concepts such as order or information. For example, negentropy in crystals with redundant order (Diamond) is not comparable to that of crystals with non-redundant order (DNA, RNA, Proteins,...). The same is true for the information required to describe a random sequence of words vs. that required for an ordered sequence of words. The concept of Redundant Information [14] helps to refine descriptions using information or negentropy. This allows us to compare the information in a morphological body plan to that in instructions in a computer program. In Information theory, redundancy measures the fractional difference between the entropy of an ensemble, and its maximum possible value. There are of course simpler ways to estimate it, as intuitively it refers to repetitions.

Another difficulty in quantifying information using negentropy is the dynamics between the information implicit in boundary conditions and free energy. Changes in this information coupled to changes in free energy are parallel to changes in negentropy, but are not identical. Information in books, objects, DNA, networks and individual and social memory are coded differently with possible different relationships between negentropy and information. Crucially, the environment is a boundary condition (information) that determines the possibilities of action [15]. The ability to produce work requires, in addition to classical forms of information, adaptive interaction with the environment. These aspects of physics are still poorly developed.

A solution to these constraints in our knowledge of physics is to learn from biology where information can be inferred from the adaptive potential of the system. Life is using information to adapt to boundary conditions using natural selection and storage of experiences, modifying the boundary conditions to increase the synergies of the system to increase the capacity of work. That is, an organism that is able to survive in more diverse and complex environments must handle more and better information. Therefore, until we get a better understanding between information and entropy, we will have to conform to the nebulous concept of negentropy and make sure we use the same type of negentropy when comparing different aspects of a complex system.

### *4.3 Manifold Order of Complexity*

As discussed when analyzing the meaning of freedom, reality has many dimensions. When analyzing open systems, we have to consider that they might be open in some dimensions and closed in others. They might be dissipative in terms of energy, for example, but in equilibrium in terms of mass or information. Many systems might be open only on certain points in time and space and in complete equilibrium in others. For example, an egg, a grain of pollen, or a seed, might be completely inactive for days, years or centuries and become active and start growing into a complex organism when receiving certain trigger signals from the environment.

This multidimensional nature of reality also occurs at the levels of natural selection. Natural selection motoring biological evolution might be acting at the level of genes, genomes, cells, individuals, families (kin-selection), closed groups (expanded family and symbioses), open groups (cultures and religions for example), diffuse groups (homophily or assortation), races, species, ecosystems, planets and other as yet undefined entities. To understand the balance between free energy and negentropy, we must make sure that we grasp the right system, the correct universe of information and the right time window of the phenomenon under study.

The view that a fine balance drives evolution of free energy or synergy has been referred to as living at "The Edge of Chaos" (coined in 1988 by Norman Packard, born in Montana 1954). For example in politics, walking the edge of chaos would imply avoiding extreme socialist views and extreme individualistic views, recognizing that the conflict group vs individual is not a problem but a driving force. Another example is steering a ship, continuous left-right maneuvers keep it on course. In innovation, the equilibrium on the edge would imply continuous changes so as to improve chances for successes above random, without increasing randomness dissipating energy.

> In ecosystems of agents, organisms, institutions, organizations, companies, networks and societies, multilevel selection is ubiquitous. Phenomenons and system dynamics in these complex settings can only be understood with a broad view from interdisciplinary sciences with a long term time frame, so as to grasp the working of creative destruction and synergistic creation on agents, firms and other entities.

Nature teaches us how to achieve synergy. This is especially revealing when comparing the biological dynamics of survival-cooperation-kin-selection-group-selection-synergy with the economic dynamics of specialization-division-of-labor-grouping-companies-democratic-decision making. [16]

### *4.4 Temporal Synchronization*

Synchrony makes possible cooperative actions in time and space, allowing coordination that is otherwise impossible. Thus, synchronous coordination achieves useful work where without it, it would simply not be economically feasible. This is possible due to two-way communications between the intervening parts. Allowing for fine coordination of the required action of each participant. This is the root of the enormous success of the cab calling application by Uber, which runs a world wide transportation network from San Francisco. The software runs from the cloud and produces synergy by synchronizing the needs of passenger and driver, smoothly coordinating the time and space of cab driver and potential passenger at the beginning and end of the ride. This synchronization produces a continuous stream of passengers for the cab and a cheap, fast and convenient pickup for the passenger. The system works because the passenger informs the system about his location at the moment he requests a car. At the same time, drivers inform the system about their location and availability to accept passengers. With this information, the systems identify the nearest drivers for any passenger request. This can be done even for drivers who are finishing, or about to finish a ride. The result of this synchronization of information is an increased overall efficiency [17] with reduced waiting time for passengers and for drivers, reducing the costs in time and money, and increasing security for both. A remarkable synergy [18] !

A non-anthropocentric example can be found among social insects. Ants communicate when retrieving food to synchronize their efforts and coordinating their efforts to drag a cockroach to their nest. They use chemical communication, secreting odorous substances on the ground which orient nestmates. Yet these odors evaporate and eventually disappear, requiring new odor depositions from recruiting scout ants. The overall effect is a synchronized recruitment of a large batch of workers to a food source that would be impossible to transport by one or a few ants alone.

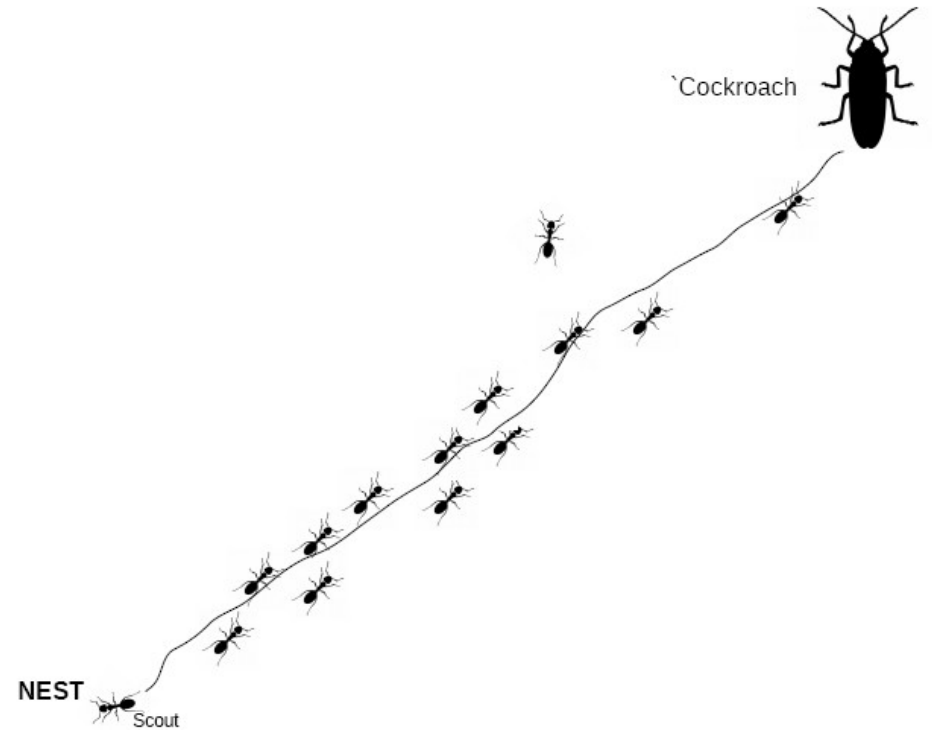

**Figure 4.4**

Chemical mass recruitment in ants. (drawing by the author )

The recruitment process starts when a scout ant finds a new food source which is too big for her to retrieve alone. She runs back to the nest, secreting a substance from her back that produces a short-lived odor trail leading from the food source to the nest. Once in the nest, the odor excites and attracts many nestmates which will follow the odor trail to the newly found food source. Once there, they collect food, and many, on their way back, will secretes fresh odors on the trail. Thus, so long as food is available, the odor trail will be reinforced. This cooperative recruitment process makes available to the ant colony a much larger range of foods than otherwise possible[19].

In both cases, synchronization achieves a nonlinear improvement of the efficiency of a specific activity. The task of synchronization requires management of richer information (more negentropy) and produces increased labor output (more free energy) thus qualifying as a synergistic process. This synergistic interaction between individuals, society and elements in the environment is often called stigmergy. The emergence of ordered dynamics from the interaction of many individuals is called self-organization.

## 4.5 Complementarity

### The membrane

Sometimes, synergy is achieved by filtering noise, blocking energy or nutrients, or defending an organism against microbes that might kill or harm. Here, filters are required. Filters might be built with different building-blocks. In Figure two examples of filters using two building-blocks are shown.

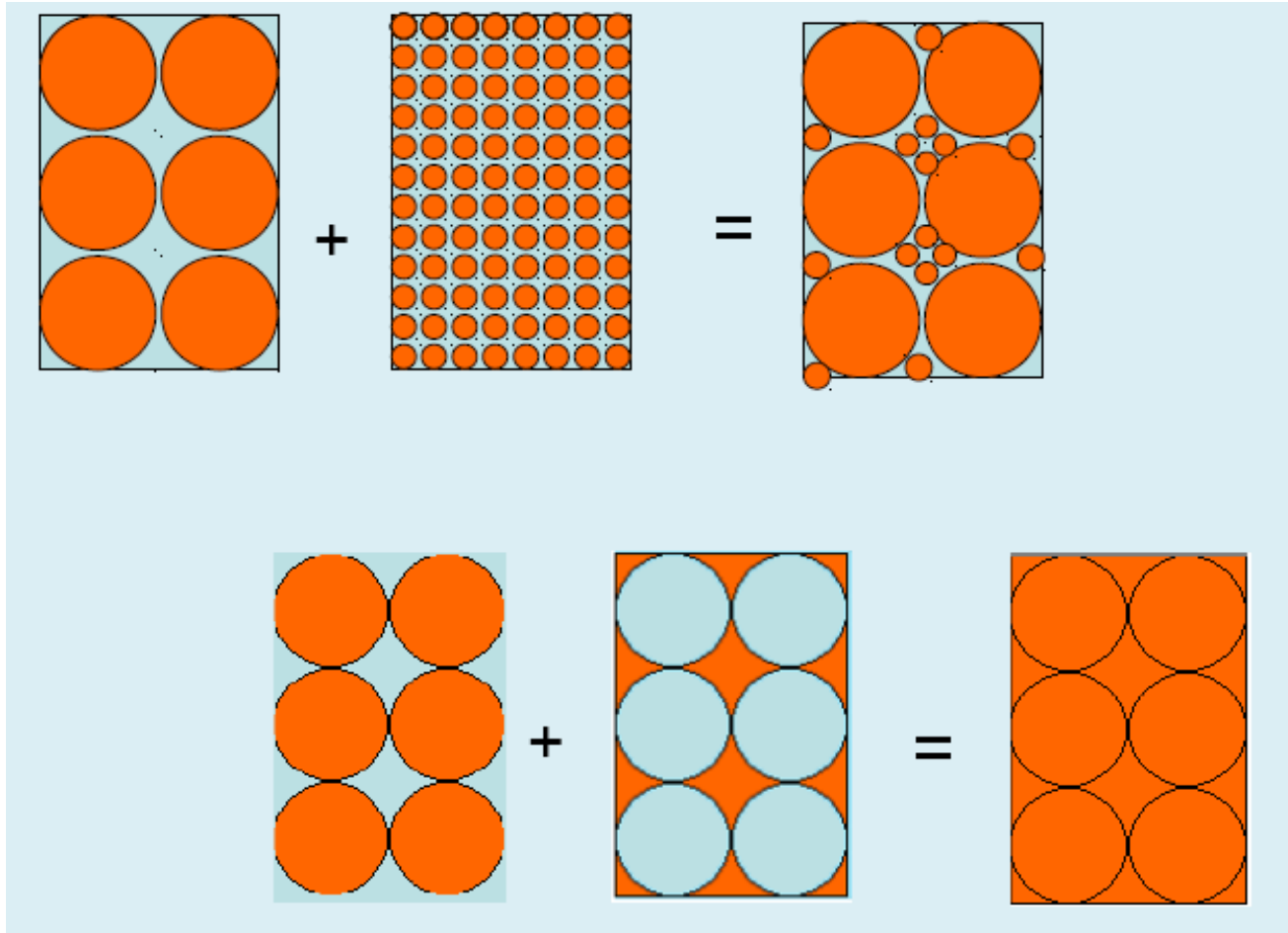

**Figure 4.5**

Building a membrane or filter. (drawing by the author )

The figure shows two examples in which different geometric forms, when pooled, produce different outcomes regarding the efficiency in filling space. If efficiency in filling space is assessed as the amount of blue area left, it is only the combination of forms in the example in the bottom of the Figure that fill the whole space, whereas in the combination at the top of the figure reduce the amount of blue space but do not fill it completely. That is, different forms interact with different efficiencies in this example regarding the filling of the blue space. Yet filling completely the blue space eliminates the blue and changes the complexity of the system, producing a new system that is qualitatively different from the first.

The phenomena allowing for efficient filtering or blocking in this example is called **complementarity**. Here, additive properties produce emergent phenomena that change the system qualitatively. In this nonlinear context, interaction of complementary forces might produce synergy. In the example, synergy seems to be negative as it reduces the complexity of the system. Cells, organisms, factories, countries and many open stable systems need membranes or borders to survive. In ecosystems, physical borders define its characteristics. If geographic accidents block the inflow of water, for example, by creating a water stopper on some critical points, some ecosystems may convert from a wetland to a dry space, allowing for example a swamp to be converted into agricultural land, or a lake in a plain. Here the synergistic effect that might be negative in relation to the flux of water has positive effects in relation to the use of land by terrestrial organisms such as humans.

This example shows that synergy has a relative aspect to it, which is part of our understanding of complex open systems following non-equilibrium thermodynamics, where a system can decrease its entropy by increasing the entropy of its surroundings. Here again, decrease in entropy or increase in order depends on the definition of the open system and where we draw the borders.

This relativity of defining the systems that reveal synergy might render the issue apt for charlatanry. Yet open systems are real entities and are easily defined intuitively. All living organisms are open systems and so are planets, stars and galaxies. Lakes and forests are open systems and despite their diffuse borders we recognize them as such. Intuition leads us where rational knowledge is lacking, but practical empirical solutions are the best. A system that cannot be quantified and measured is useless for scientific and engineering purposes. It should be left as fodder for charlatans.

**Nest guarding by wasps**

A simple and important example of complementarity can be found in wasp's parental investment. In many wasp species, females lay eggs in specially built nests. There the offspring hatches as larvae and are fed by the mother until the larvae metamorphose into an adult wasp. During that time, the female must both stay in the nest to guard the brood and defend it against predators (other wasps and spiders), and forage for food (insects) to feed the larvae. Both activities are incompatible. Thus, the odds of a female wasp to rear successfully their broad to adulthood is practically zero. Yet, if the females cooperates with other wasp females in rearing their offspring, complementing their activities so that when some forage the others guard the nest and vice-versa, a synergistic effect ensures. Survival of offspring is much more than the addition of survivor-ship of each female; it becomes practically 100%.

A similar situation has been reported among humans. The survival rates of fetuses and of children increase when the mother profits from cooperation from relatives and/or a spouse.

**Innovation and equilibrium**

Innovative cultures “must be counterbalanced by some tougher and frankly less fun behaviors. A tolerance for failure requires an intolerance for incompetence. A willingness to experiment requires rigorous discipline. Psychological safety requires comfort with brutal candor. Collaboration must be balanced with individual accountability. And flatness requires strong leadership. Innovative cultures are paradoxical. Unless the tensions created by this paradox are carefully managed, attempts to create an innovative culture will fail." [20]. This very pertinent insight by Gary P. Pisano (born 1961) is a reflection of how synergy is achieved in innovative enterprises.

Generally speaking, complementarity can work as a balancing mechanism achieving equilibrium at the edge of chaos. The term edge of chaos, i.e. a transition space between order and disorder that is hypothesized to exist within a wide variety of systems. This transition zone is a region of bounded instability that engenders a constant dynamic interplay between order and disorder. [21] Its precise nature is elusive to grasp as it relates to areas of knowledge we are still discovering and expanding.

## 4.6 Biological Symbioses

The most famous examples of synergy through complementarity are found among the large number of symbioses known to exist in nature. The term "symbiosis" is generally used by biologists to connote the "living together" of "dissimilar" organisms for their mutual benefit. Here I will mention just a few. Many more examples have been published by P.A. Corning[22], which include other processes in addition to symbioses, which I highly recommend reading.

### Lichens

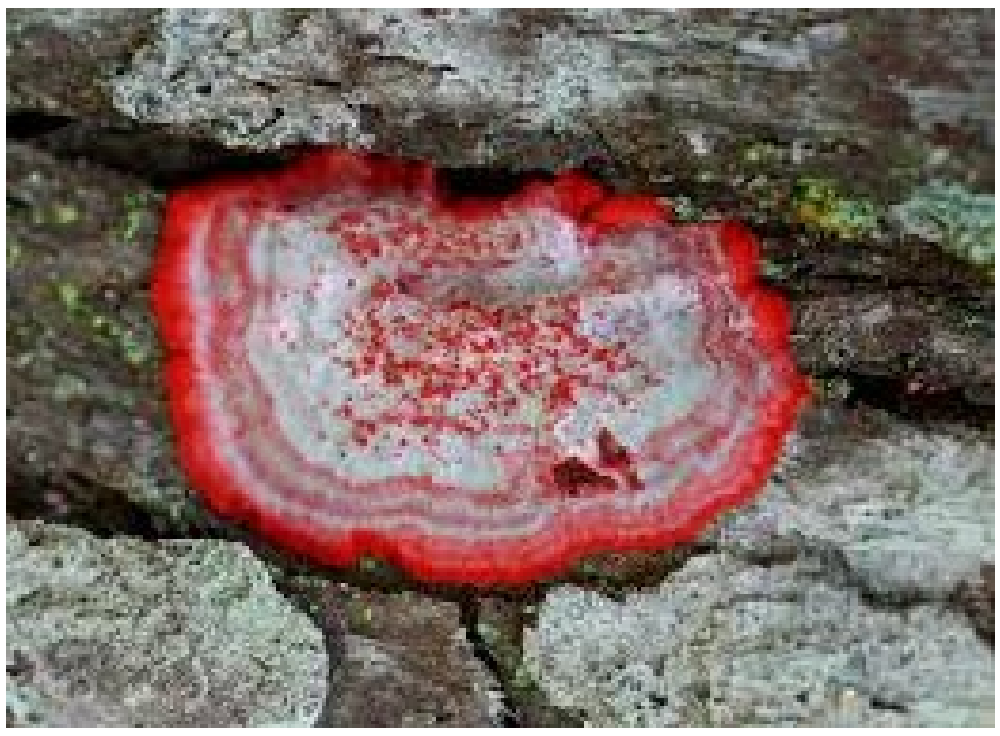

**Figure 4.6a**

Lichen growing on rocks. Source: Google Images

Lichen are the product of a mutualistic symbiosis, forming around roughly 20,000 different species of partnerships between some 300 genera of fungi and various species of cyanobacteria and/or green algae. The combined lichen has properties different from those of its component organisms. The properties are sometimes plant-like, but lichens are not plants. In this partnership, the fungal part helps capture nutrients, whereas the alga captures energy from light through photosynthesis. Although many lichen partners can apparently live independently, in combination they enjoy significant functional advantages. They are the organisms that can adapt to the most extreme environments ranging from Antarctica to the Tropics and from sea level to high up the glaciers and deserts of mountains.

**Agriculture**

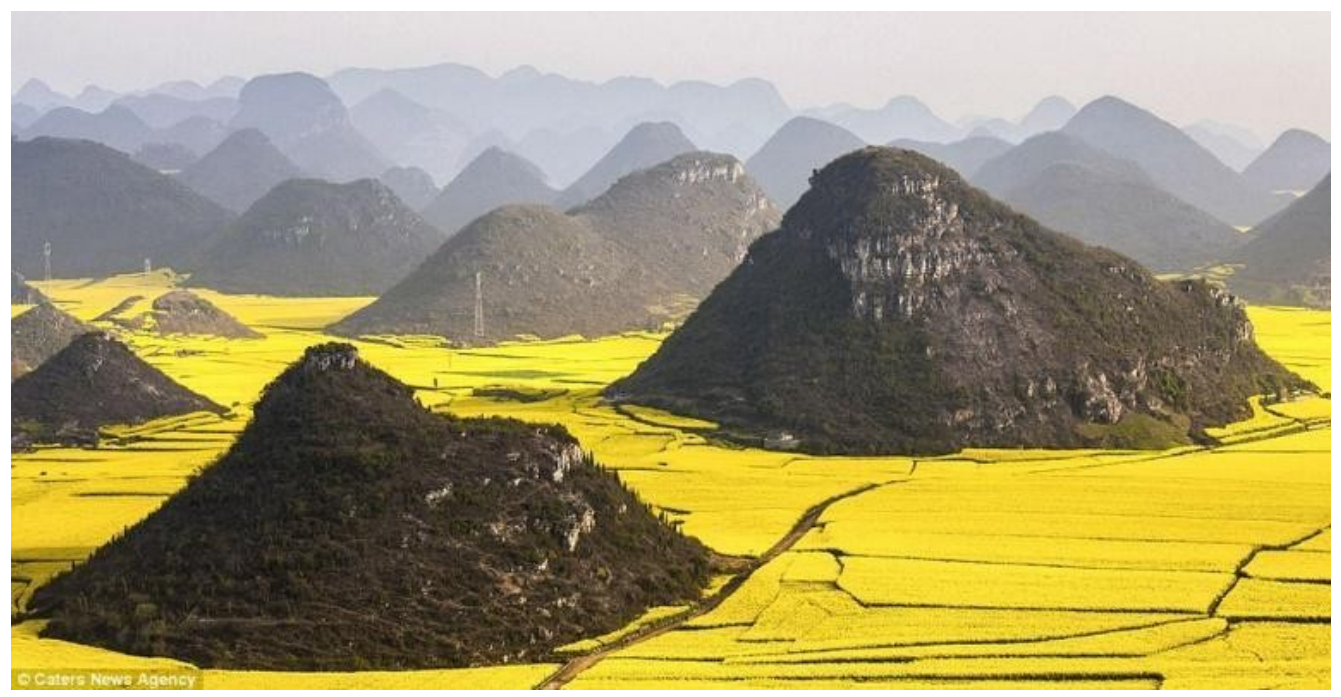

**Figure 4.6b**

Extensive oil seed plantations in China. Source: Google Images

Leaf-cutter ants collect freshly cut leaves to their nests to feed cultures of specially domesticated fungi on which they and their larvae feed upon. They have developed special mechanical and chemical procedures to keep the cultures healthy and block infections by other organisms. They provide the ideal temperature and moisture for an optimal growth of their fungal cultures. These fungi cannot live without their ant hosts, nor can the ants live without their fungal cultures. The fungus destroys insecticides and toxins accumulated by the plant in the leaves to avoid herbivory. It is a true symbiosis.

Humans have domesticated a range of plants and animals. In that process, these plants and animals have become so dependent on their human hosts that most of them are now unable to thrive in nature without the help of their host. Humans, on the other hand, have become so dependent on their domestic plants and animals that a humanity based on collecting and chasing free living plants and animals is completely impossible. Both parts of the symbiosis are benefiting so much from this partnership that neither can live without the other. This constitutes a truly synergistic relationship.

**Microbiome**

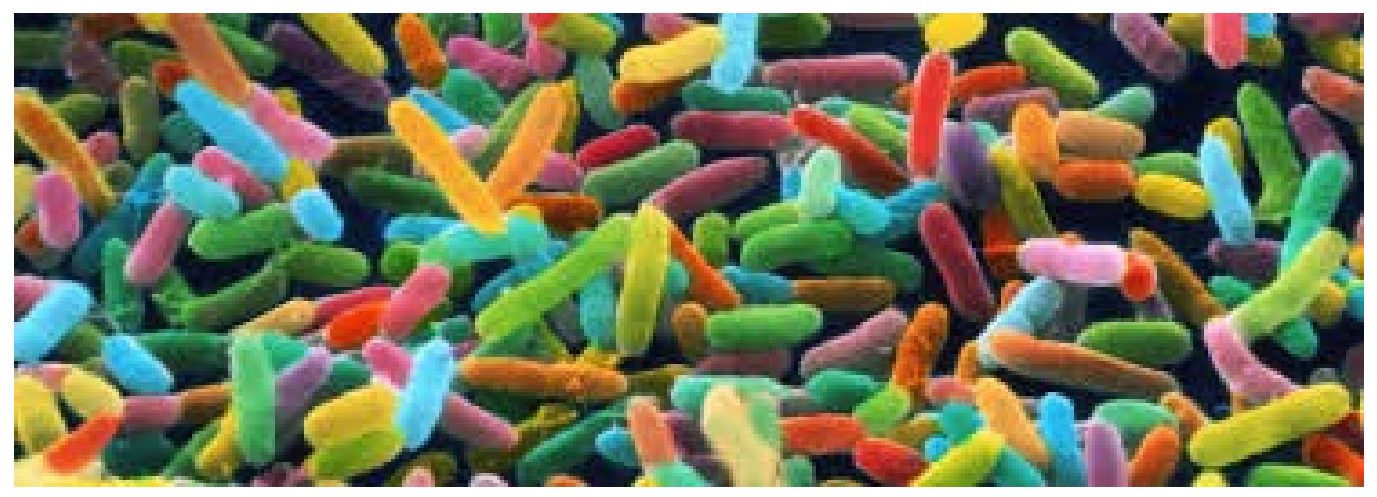

**Figure 4.6c**

Bacterial culture: Source: Google Images

Many microorganisms are found in association with both healthy and diseased humans. New evidence of the importance of microorganisms for our nutrition and health are forthcoming every year. We know that the microbiome determines our risks for obesity and diseases. Several authors even claim the number of bacterial cells in our body matches and even surpass that of our own cells. Clearly, humans and most animals are not able to survive without their symbiotic microorganisms. The generic term bacteria should be substituted for that of microbiome, as it refers to bacteria, Archaea, yeasts, and single-celled eukaryotes, as well as helminth parasites and viruses. Some surprising finds about the human microbiome include:

- Microbes contribute more genes responsible for human survival than humans' own genes. It is estimated that bacterial protein-coding genes are 360 times more abundant than human genes.

- Microbial metabolic activities; for example, digestion of fats; are provided by several bacterial species.

- Components of the human microbiome change over time, affected by a patient disease state and

medication. However, the microbiome eventually returns to a state of equilibrium, even though the composition of bacterial types has changed.

Plenty remains to uncover about the symbioses with our microbiome.

# 5. Social Synergy

## *5.1 Definitions*

Synergy in politics, in social settings, in social networks, in relationships, etc. is ubiquitous. Many successful politicians promote reciprocal positive feedback among networks of minority voters that guarantee a winning outcome in elections. Political parties build social support that allows them to surpass the threshold needed to gain power. Sustainable policies aim at win-win outcomes for users and suppliers of public services. Humans submitted to poverty and nutritional stress benefit from synergistic networking making them less vulnerable and more resilient. Technological societies successfully manage knowledge, trust, wealth and human resources to achieve sustained economic growth.

Many social philosophers in the past and present have claimed, in one form or another, that the value of society is much greater than the aggregate values added by individuals. In some form or another, they have proposed specific relationships between individuals to form the basis of the whole system. Today there seems to be a consensus that cooperation between individuals is the most important relationship in creating social value. This cooperation is the root of the **Social Contract** or the voluntary agreement among individuals by which organized society is brought into being and invested with the right to secure mutual protection and welfare or to regulate the relations among its members. This rationale has been promoted by many philosophers, including Thomas Hobbes (born 1588 in Westport; died in 1679 in Derbyshire), John Locke (born 1632 in Wrington; died in 1704 in High Laver, UK), and Jean Jacques Rousseau (born 1712 in Geneva; died 1778 in Ermenonville, France), Many specific and detailed proposals have been made over the years, explaining how this cooperation produces social value. A few examples are given here:

- **Altruism** is proposed by many religions as fundamental for the health of society. The mostly unspoken rationale behind this thought is that the value of a gift exceeds that of the cost to the donor. For example, an old spare blanket that might just occupy valuable space in a home, if donated to a freezing homeless person on the street, will be of much more value to that person than to the donor. In this way, charitable donations increase the overall good of society. The limitations of using altruism as an explanation for the stability of society is that in the long run, a pure altruism is not viable, as it reduces the chances of survival (fitness) of the altruist. Pure altruism cannot evolve through biological nor through cultural evolution. Something else is needed.

- **Mutualism** or the doctrine that mutual dependence is necessary to social well-being was advocated by the philosopher Pierre Joseph Proudhon (born 1809 in Besancon; died 1865 in Paris). Mutual aid is beneficial to all participants and forms social bonds that cement society. Mutual bonds can be based on cooperation, solidarity, mutual defense and/or commerce. Without these interactions, societies would not exist. The shortcomings of mutualistic explanations for social evolution is that free-riders, parasites and cheaters will eventually gain the upper hand making long term mutualistic relationships unlikely. Several limits to mutualism have been proposed. The most important is greed and short-term selfishness. The **prisoner's dilemma** is a standard example of a game analyzed in game theory that shows the rationale behind why individuals might not cooperate, even if it appears that it is in their best interests to do so. It was originally framed by Merrill Flood (born 1908 in Seward and died 1992 in La Jolla) and Melvin Dresher (born 1911 in Poland; died 1992 in Bakersfield) working at RAND in 1950. In this game of two prisoners awaiting a trial, the player that defects, denouncing the companion to be himself freed, benefits at less risk than a player who remains silent in the hope that the companion will cooperate by also not speaking, even though cooperation would yield higher overall benefits to both prisoners. The same dilemma appears in what economists call "The tragedy of the Commons". A forest, grazing ground, or a fishery that has no owner, be it a communal owner or an individual, is bound to be exhausted as individuals have no incentive to show restraint in overexploiting the resource, because what you don't take, another will take.

- **Reciprocity** or the rule that you should pay in kind what another person has provided to you was proposed as the building block for society by sociobiologist Robert Trivers (born 1943 in Washington) and others. Specifically, Trivers defined **reciprocal altruism** as a behavior whereby an organism acts in a manner that temporarily reduces its benefits while increasing another organism's gains, with the expectation that the other organism will eventually act in a similar manner later. In order for reciprocity to work, free-riders, parasites and cheaters have to be identified and isolated. This is not always feasible.

- **Conditional reciprocity** was explored by the political scientist Robert Axelrod (born 1943 in USA) inducing the discovery by Anatol Rapoport (born 1911 in Lozova, Ukraine and died 2007 in Toronto) of a very successful strategy know widely known as "Tit for Tat", or "blow for blow" and "help for help. Here, an individual engages in partnerships only with individuals which also cooperate, avoiding the costs of interactions with defectors. If you can't know in advance if a possible partner will cooperate or not, a second-best strategy is to simply cooperate if your former partner cooperated, and defect if the former partner defected. This strategy is called Tit for Tat.

More sophisticated strategies do not add much efficiency to his one.

- **Indirect reciprocity**, promoted by the biologist Richard D. Alexander (born 1925 in Piatt County, USA), and in a different form by the economist Robert Frank (born 1945 in Coral Gables, Fl), assumes that **reputation** will add value to an individual in future cooperative encounters by increasing the trust individuals in his society will place on him, facilitating reciprocity and cooperation in social interactions. This works because cooperative individuals gain a reputation that helps them to find better cooperative partners in the future. Reputation, however, might also be faked. Thus, reputation and reciprocity on their own do not seem sufficient to explain social evolution.

- **Altruistic punishment** is another mechanism that has been proposed to nudge the social contract. If free riders, parasites, defectors and greedy individuals escape cooperation for selfish reasons, increasing the cost of defection will reduce the likelihood of defectors by reducing or eliminating their gains. Punishment, however, has cost for both, the punisher and the punished, and complex possible webs of interactions can emerge.

- The concept of the **selfish genes** and of Inclusive Fitness, as developed in Sociobiology, overcomes many of the limitations stated above. That is, cheating against oneself does not make sense. The insight here is that in evolutionary biological speak, the self is not the organism, but the gene, or pool of genes (or memes). If a gene codes for an altruistic or mutualistic act, its effect will be suffered eventually by many organisms carrying that gene. Altruism towards relatives and carriers of some of our genes benefit those genes. This is now called Inclusive Fitness and will be explored in the next chapters.

### 5.2 Synergy from Cooperation

All cooperative encounters have costs, but defection is not always profitable. In many cases, cooperation is a win-win strategy for all involved. Synergy, or a multiplier effect, shifts the cost/benefit balance of free-riding and cooperation sharply towards cooperation. The biologist Edward Osborne Wilson (born 1929, Birmingham, Al) described "The **multiplier effect**"[23] as follows: "*In the context of organizational behaviour is the view that a cohesive group is more than the sum of its parts. Synergy is the ability of a group to outperform even its best individual member*". Clearly, society is expected to produce synergies. A society without synergies is unlikely to be stable. Social synergy is the backbone of animal and human societies.

Computer experiments have been used to explore the logical rigor of each of the propositions listed above for the working of social cooperation. Using advanced simulations of artificial societies, it could be shown that none of the proposed features listed above can explain the emergence and maintenance of social behavior. The arithmetic bean counting of the prisoner's dilemma choices do not add up in complex virtual societies that mimic real ones. Nor are there magical genetic systems that make social behavior more likely to evolve. The only single feature that can produce viable and stable societies is the inclusion of a synergistic effect in cooperation. The following simple examples will explain what synergistic cooperation means.

- A single stone age hunter is able perhaps to hunt on its own a bird, a frog or a rodent. Two might hunt a deer and transport it to their family, but a dozen might kill and butcher a mammoth, with over a ton of meat, whereas any of the smaller prey stone age hunters could retrieve on their own would weigh a few Kilograms. Thus, cooperation among stone age hunters produced synergies that benefited each of the participating individuals in addition to large benefits to society. No prisoner's dilemma reasoning was needed here.

- A lone fisher in the Middle Age could catch several small fish and even a big one on its own by using the fishing tools available. A small group of whale fishers, however, could fish one or several adult whales, which represents tens of thousands of times more meat than the average lone fisher. Again, the benefit each individual fisher received for fishing cooperatively, by large exceeded what he could expect from fishing alone.

- A XXI Century pensioner lives alone in his house. Every morning he sweeps the front of his house, and by the way, cleans that of his neighbors. Because of this behavior, the neighbors regard him as an altruist. When one day, a heart infarct strikes him, he has just enough strength to call a neighbor, who takes him to hospital, saving his life. The value the old man assigns to its life, of course, is infinite. Here is no amount of work, money or time that can buy a new life for him. The time and effort spent in attending the neighbor's street front is a small change compared to what he values in his life. For him, his acts are not altruism, they are a very cost-effective way of social investment.

The amount of synergy unleashed in this example is very large. If we assume groups of 20 individuals, and average individual catches of lone hunters of 10 Kilograms, the social synergy value you must use to multiply the average benefit of solitary individuals to get the average benefit to individuals in the cooperating group is 5 in the first example, 50 in the second, and infinitely large in the third. Computer simulations show that stable societies emerge in evolution with even smaller values for social synergy. In the virtual worlds, societies need only values of 2, or even of 1.2, to increase the odds for the emergence of social cooperation above 50%. Values of social synergy of 3 or more produce societies based on cooperation with odds of 100%. Achieving the social benefits with smaller groups increases this synergistic effect. Thus, extrapolating the social simulation results to real life makes the emergence and maintenance of social cooperation unavoidable.

Alfred Emerson (born 1896 in Ithaca, NY and died 1976 in Lake George NY) proposed [24] that evolution is much more concerned with cooperation than with competition. An empirical example can be found in the evolution of mutualism between butterflies and ants. Butterflies larvae of species in the butterfly family Lycaenidae have special relationships with ants. Ants protect the butterfly larvae from predators and larvae provide nutrients to the ants. Based on the types of interactions between lycaenid larvae and ants, the myrmecophilous organs on the lycaenid larvae, the degree of relationship (facultative or obligate) and the diet of the larvae, it can be shown [25] that cooperative symbiotic interactions between ants and butterfly larva, where both parties benefit from the interaction, have appeared much more often in evolution than exploitative relationships, where one part is considered a parasite of the other. This shows empirically the stability of cooperative symbiotic interactions in evolutionary terms.

### *5.3 The adaptive value of Love*

Computer experiments that mirror the evolutionary dynamics of sexual and asexual organisms as they occur in nature, tested features proposed to explain the evolution of sexual recombination [26]. Results show that this evolution is better described as a network of interactions between possible sexual forms, including diploidy, thelytoky, facultative sex, assortation, bisexuality, and division of labor between the sexes, rather than a simple transition from parthenogenesis to sexual recombination. Diploidy was shown to be fundamental for the evolution of sex; bisexual reproduction emerged only among anisogamic diploids with a synergistic division of reproductive labor; and facultative sex was more likely to evolve among haploids practicing assortative mating.

The main conclusion of this research is that the emergence and maintenance of sexual reproduction among living organisms is not possible if the interaction between the parents does not include some kind of synergy that benefits directly or indirectly the offspring. Thus, rather than looking at natural selection as a conflict between sexes, as many a biologist does, we need to focus on understanding the synergies that emerge among the participants of sexual reproduction.

For example, the appropriate mixing of genes achieved through assortative mating may confer increased immunity against pathogens to the offspring, or complementary skills inherited by offspring may confer novel benefits for survival. On the parents' site, cooperation in parental investment, rearing of the brood, construction of nest, defense of offspring, etc, my proof to be synergistic, as the example of the wasps showed (see section 4.5).

This fundamental synergistic arrangement at the origin of sex inspired the title of “Information and Energy: Gods of Genesis”, suggesting that synergistic coupling of two entities is at the roots of most synergies. More than two sexes, although possible, is unlikely [27].

## 5.4 Enterprise: Wealth from Synergy

The activity where professionals are most aware of the workings of synergy is probably economics. However, economists do not consider wealth as a free energy and money is mainly considered as a lubricant for executing economic transactions. This insight is rather misleading. None of the main general economic theories speak about synergy explicitly. It is present very much when planning mergers of companies, and it is inferred when considering groups of cooperating workers. The economist Adam Smith (born 1723 Kirkcaldy and died 1790 in Edinburgh) coined the concept of the invisible hand of markets and showed, for example, how a group of specialized workers in a pin factory could produce together much more efficiently and a much larger quantity of pins than what in aggregate the same workers could produce if working each one on their own. But not all groups of workers outperform individuals working alone. A squabbling group of merchants will be much less efficient in their business than an organized rich merchant doing business on its own.

Adam Smith in his book The Wealth of Nations

describes what is by now a very famous example. "*… the trade of the pin-maker; a workman not educated to this business (which the division of labour has rendered a distinct trade), nor acquainted with the use of the machinery employed in it (to the invention of which the same division of labour has probably given occasion), could scarce, perhaps, with his utmost industry, make one pin in a day, and certainly could not make twenty. But in the way in which this business is now carried on, not only the whole work is a peculiar trade, but it is divided into a number of branches, of which the greater part are likewise peculiar trades. One man draws out the wire, another straights it, a third cuts it, a fourth points it, a fifth grinds it at the top for receiving the head; to make the head requires two or three distinct operations; to put it on, is a peculiar business, to whiten the pins is another; it is even a trade by itself to put them into the paper; and the important business of making a pin is, in this manner, divided into about eighteen distinct operations, which, in some manufactories, are all performed by distinct hands, though in others the same man will sometimes perform two or three of them. I have seen a small manufactory of this kind where ten men only were employed, and where some of them consequently performed two or three distinct operations. But though they* *were very poor, and therefore but indifferently accommodated with the necessary machinery, they could, when they exerted themselves, make among them about twelve pounds of pins in a day. There are in a pound upwards of four thousand pins of a middling size. Those ten persons, therefore, could make among them upwards of forty-eight thousand pins in a day. Each person, therefore, making a tenth part of forty-eight thousand pins, might be considered as making four thousand eight hundred pins in a day. But if they had all wrought separately and independently, and without any of them having been educated to this peculiar business, they certainly could not each of them have made twenty, perhaps not one pin in a day; that is, certainly, not the two hundred and fortieth, perhaps not the four thousand eight hundredth part of what they are at present capable of performing, in consequence of a proper division and combination of their different operations.*"

At the roots of this description is economic synergy, although Adam Smith did not use this word. Not all economic activities, however, produce synergy. In the Figure a schematic summary of possible economic social interactions is represented.

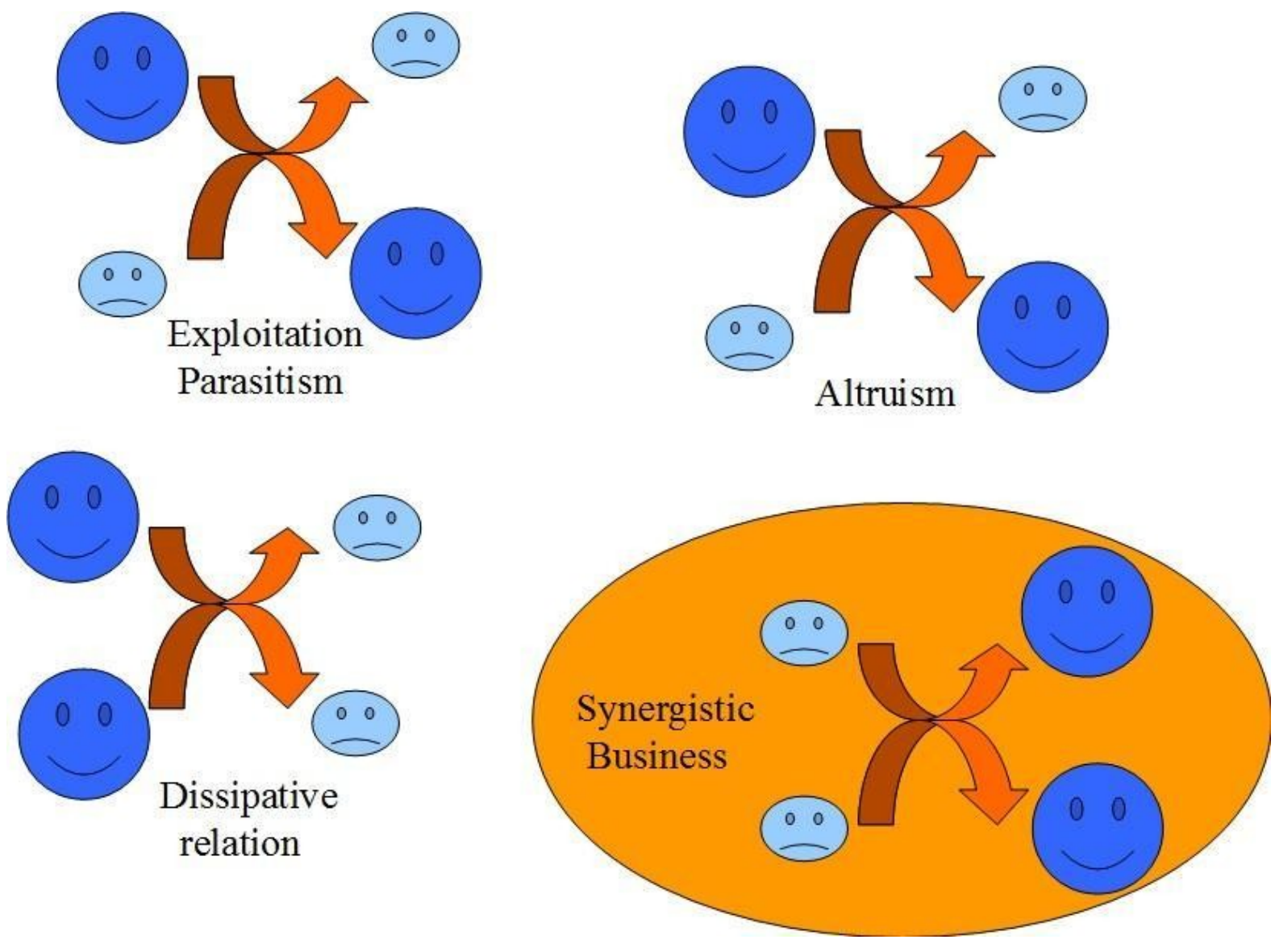

**Figure 5.4**

Types of economic interactions (drawing by the author)

In pure altruistic interactions, one of the partners in the economic exchange reduces its wealth, whereas the other partner increases it. This is the same outcome from an exploitative or parasitic interaction. The difference between both descriptions is semantic: in one case, we describe the part that reduces its wealth as an altruist; in the other case, we call it the victim of exploitation. The difference in economic terms between parasitism and altruism is nil. In psychological and emotional terms, there might be many differences, but the impact over the aggregate accumulated wealth is equivalent.

In a synergistic interaction, aggregate wealth increases. Many altruistic acts are really, in economic terms, a social investment. When a wealthy donor gives a poor recipient a blanket, the recipient will get a much higher utility from the blanket than the donor, but there is no net increase in wealth. But if the object donated is a sewing machine, which is used in the rich donor's house as decoration, but the poor receiver uses it to produce blankets to sell, then the increased utility might be transformed into increased wealth. It is this second type of synergistic interaction that is synergistic in economic terms. In economic terms, it should be called a social long-term investment rather than an altruistic act, as all participants will eventually benefit from this investment. Less poor street dwellers will indirectly also increase the quality of living conditions of a wealthy donor.

Frequently, economic exchanges dissipate aggregate wealth and all participants in the interaction lose. This is often the case of individual altruistic punishment.

Here, an altruistic individual punishes a transgressor of a norm at a cost to himself and to the punished individual. When enforcement is random, the cost to the punished is low, and the benefits of transgressing the norm high, the resulting economic outcome is dissipation of resources. This favors punishment through a centralized authority which is much more efficient in enforcing social rules when punishment is costly. This is one of the reasons for the emergence of centralizing state bureaucracies. They can train and maintain specialized repressive forces to provide the required punishment efficiently.

The division of labor is the single most important feature that enables synergistic interactions as it allows for simultaneous execution of tasks that then can be synchronized. It is the most conspicuous element in an efficient market. Synchronizing labor requires trust. Efficient division of labor requires specialization, and specialization requires knowledge. Thus, the roots of synchronization and division of labor produce knowledge and trust, which are always part of the prosperity syndrome[28] in economics. It is the presence of social synergy achieved through division of labor that allows markets to be efficient in achieving wealth and health in a society.

The reason why division of labor is essential for a synergistic economic process can be illustrated in a simple example of world containing only two resources, let's say sugar and salt[29]. Omnipotent agents inhabiting this world can collect sugar and salt. They can exchange it with others, to make profits in each exchange, increasing their personal wealth. Alternatively, agents can be specialized salt miners, sugar farmers or traders, limiting their activities to only one of the possibilities an omnipotent agent has. A fast and intuitive assessment of this situation would think that omnipotent agents will be better off in aggregate terms compared to specialists, as they are less likely to be economically inactive due to their broad possibility of action. Computer experiments using this imaginary world show that this intuitive assessment is true only if both resources are evenly and uniformly distributed. If the distribution of these two resources, however, is heterogeneous, with clumps of salt mines in one place and areas with sugar farmers somewhere else, specialists will always do better than omnipotent agents in aggregate terms. That is, synergistic interactions through division of labor are favored by the heterogeneity of the environment. The more complex and diverse the environment, the more specialization of a task improves performance, and the more division of labor is needed for synergistic economic processes[30].

The real word, of course, is much more complex than the example presented above. True omnipotent economic agents are not likely to exist and the capabilities of individuals to dominate the knowledge, training and abilities required to execute all possible tasks in an economy is nil. Thus, complex economies require division of labor, producing synergies whereby all participants benefit, producing win-win interactions that favor and stabilize social arrangements that favor them. This is not true in very primitive hunter-gatherer societies. Possibilities of division of labor are fewer and zero-sum economic interactions are more likely. If someone fishes all the fish in a lake, there will be nothing left for other fishers. If however, individuals cooperate in planting a corn field, or in producing pins, they can organize themselves so as to optimize the different activities, unleashing synergies that benefit all. These examples give us a glimpse into the working of the invisible hand of the market.

### *5.5 Synergy in Business*

The human activity where consciously or unconsciously we are most interested in achieving synergy in the processes we manage is business. Some concrete examples might throw light on how synergy might be achieved in business. Specifically, as shown above, size often favors the emergence of synergistic interactions. This is not always the case in business.

Here a few examples:

1. Acquisitions, in general, have been demonstrated to create economic value. The intuitive reason underlying this value creation stems either from an ability to reduce costs of the combined entity, an ability to charge higher prices, or both. Current research in the area attributes these abilities to an opportunity to utilize a specialized resource. A study compared three broad classes of resources that contribute to the creation of value[31]. Following the conventional wisdom, these resources are classified as cost of capital related resulting in financial synergy), cost of production related (resulting in operational synergy), and price related (resulting in collusive synergy). The study found that collusive synergy is, on average, associated with the highest value. Further, the resources behind financial synergy tend to create more value than the resources behind operational synergy.

2. Are large firms more profitable, regardless of the extend of division of labor, and is their survival rate better than smaller ones[32]? This paper uses Australian time series and cross-sectional data at the firm level for the first half of the twentieth century to address these questions. It compares the return on equity for the largest 20 companies in a database with the remaining 405 firms for the period 1901-21, and then calculates the comparative attrition rates over the following forty years. The top 20 firms were more profitable, though with somewhat mixed individual performances, and had lower attrition rates over the following 40 years than smaller firms. Thus, larger firms, on average, performed better and were more effective at sustaining their relative standing.

3. The use of artificial neural network (ANN) to explore

the nonlinear influences of firm size, profitability and employee productivity upon patent citations of firms in the US pharmaceutical industry is revealing[33]. The result shows that firm size, profitability and employee productivity have the monotonically positive influences upon patent citations of the pharmaceutical companies in the US. Therefore, if a pharmaceutical company wants to enhance its patent citations, it should enhance its firm size, profitability, and employee productivity.

4. The relationships among firm size, profitability and diversification are examined for a sample from the top 400 industrial firms in Canada in 1975[34]. Account is taken of industry-specific factors and of foreign ownership. The main findings are that increasing firm size is not associated with higher profitability, larger firms do appear to experience greater profit stability, and the relationship between firm size and diversification is positive but weak. Industry factors are far more important than firm size in determining inter-firm variations in diversification, implying that diversification is not undertaken to stabilize profits by all large firms.

5. The relationship between the corporate social performance of an organization and three variables: the size of the organization, the financial performance of the organization, and the environmental performance of the organization, also shows the existence of synergy in business[35]. By empirically testing data from 1987 to 1992, the results of the study show that a firm's corporate social performance is indeed impacted by the size of the firm, the level of profitability of the firm, and the amount of pollution emissions released by the firm.

6. Quality of interpersonal relations also matters. In a study by Alex Edmans at the London Business School of the 100 best places to work in the US he evaluates their economic success according to their working atmosphere, tracking them over 25 years. He found that the top best places to work increase their share value 50 percent relative to the others.

7. According to some electrical contractors who are members of the Federated Electrical Contractors, electrical contractor firms may experience a lack of profitability as the firm grows in size[36]. Under these conditions, statistical models were developed to study the firm's size-profitability relationship. Economic data were obtained from the National Bureau of Economic Research, Bureau of Economic Analysis, and Mortgage Information Service. Financial data for 1985–1996 were obtained from the FEC group. Statistical analysis reveals that small, medium, and large firms are significantly different from each other in terms of their profit rate; profitability drops as firms grow larger than $50 million in sales. An indicator variables model with a first-order autoregressive model built into the error term was developed using backward elimination regression. Data from the year 1996 were used to validate the model, which predicted 76% of the year 1996 response variable, profitability, correctly.

8. Mergers are not always profitable. Using data on 2,732 lines of business operated by U.S.

manufacturing corporations, a paper[37] analyzed the pre-merger profitability of acquisition targets and post-merger operating results for the years 1957–1977. Acquired companies are found to be extraordinarily profitable pre-merger, the more so, the smaller their size. Following merger, the profitability of acquired entities declined except among pooling-of-interest's merger partners of roughly equal pre-merger size. The decline was larger than expected under Galtonian regression. This and the high divestiture rate for acquired entities point toward control loss as an explanation of the profit drop.

9. Several studies show that the hypothesis for a relation between board size and financial performance does not always hold[38]. Empirical tests of the relation exist in only a few studies of large U.S. firms. We find a significant negative correlation between board size and profitability in a sample of small and midsize Finnish firms. Finding a board-size effect for a new and different class of firms affects the range of explanations for the board-size effect.

These studies show few consistent predictors for successful mergers. Clearly, not always size favors the emergence of synergistic interactions. Often, synergies are lost when companies merge or increase too much in size. Control of completion or increased monopolistic power rater than increased synergy explain many mergers.

The example hints that improved specialization, complementarity of the integrated parts of the merger, increases in available expertise, better harmonic integration of workers into the firm and increased quality of life at the workplace, are factors that relate to synergy consistently.

In the discussions about the fragility of public goods versus the powers of Leviathan, synergy is known to emerge though institutions such as private or communal property. A large body of evidence has accumulated of the working of synergy in these cases, thanks in large part to the work of the political economist Eleanor Orstrom (born 1933 in Los Angeles and died 2012 in Bloomington IN), winner of the 2009 Nobel prize in economics. Division of labor, cooperative networking and increased motivation are all ingredients for achieving synergistic outcomes in managing resources.

But the factor most correlated to synergy in business is knowledge of the scientific kind. Whatever variable was correlated with the wealth of a nation, the strongest correlation was always related with knowledge of the science [39]. That is, countries with large academic sectors engaged in basic scientific research in the natural sciences had much stronger economic growth and had more total wealth than those with less research activity. This relationship, although also present, was less strong when correlating the strength of economic institutions, technological knowledge, arts and films, and many other indices. That is, deep knowledge of the phenomena to tackle allows to avoid the working of conflicting forces and enable the system to achieve productive synergistic interactions.

In general, synergy in time and space drives economics: "Velocity" of money relates to efficiency of economy. That is, economic efficiency is affected by the speed of the activity as well as by the amount of wealth involved in the transactions. Many a synergistic gain can be achieved by diminishing the obstacles to economic activity by implementing rules and laws that organize social interactions, but beware of excessive rule writing [40].

### *5.6 A Social Contract*

Societies are a product of biological and cultural evolution. Sociobiological research has shown that a key feature in all these societies is that each and every member of the society has an important role. That is, in societies of plant cells, the role of a cell at the base of a tree trunk is as important as that of a cell at the end of a branch. In human cities, the street cleaner is important for social hygiene, and so is the medical specialist at the central hospital. Social synergy arises when a harmonious coexistence and cooperation of all parts of society is achieved so that all participants profit in some way from belonging to the society.

Among humans, the biological cement of social bonds are provided by behaviors and sentiments, molded by nurture and nature, such as trust, lust, love, respect, affection, power-seeking, shame, empathy, humor, humility, and others, which form a web of forces interacting synergistically producing long term benefits to actor often in unpredictable ways. These webs of interactions constitute a social contract. Behaviors that make transactions more efficient, relationships more mutualistic and investments more likely to produce benefits, in the short and long term, can all be considered to be synergistic regarding social interactions. These win-win strategies enrich the social contract over time.

Successful political systems not only have to handle national economies, but have to manage the social-political environment so as to allow our innate social sentiments to congregate around social synergies. A large number of institutions, invented by humans in different periods of our history have proved successful for advancing this task. **Democracy** in its various forms has been indicated more frequently in modern times as the best of these institutions. But democracy has various forms. **Populist Democracies**, which focus on very short term benefits without regard to longer term impacts of policies are often very harmful to society. **Meritocratic Democracy**, often tending to **Democratic Aristocracy**, seems to have served industrial society well. **Autocracy**, however, seems to be the default mode of institutional management when civil society has no means to stop it. Political science has many examples and lessons to teach in this regard.

A society governed by institutions achieves social regulation by balancing opposed forces. All regulation has to be implemented rationally, with continuous oversight about its working and its beneficial effect on the synergistic working of the group. Regulations often harm the working efficiency of the group, and not always because of its sloppy implementation. A key for the successful balance of divergent forces is a synchronous communication (section 3.4) which implies synchronization of the different dynamics of the diverse components of the system.

The ideological political framework that best might manage the forces of synergy is **Liberalism**. Not the dogmatic economic liberalism, but a kind of empirical scientific liberalism which accepts that social and economic progress can be achieved with less top down planning and more bottom up dynamics. The laws of biological evolution show us that diversity and liberty may work wonders; that uncertainty is a universal social law, even in physics (Schrodinger's uncertainty principle); that individuals are constrained due to many reasons, including limitation in visualizing different time horizons (from fractions of milliseconds to millions of millennia); that there are increasing difficulties in managing more distant information (homologous to the law of gravity in physics); that social Investment (i.e. altruism) and the biological dynamics in social interactions partially determine social and economic progress of human communities. These features are indispensable when planning human interventions is social regulations.

An important element for a liberal democracy to work is **Empathy.** Empathy is the tool that knit social synergy, whereas Greed is the stupid attitude that bypasses empathy: Enlightened self-interest is clever but works only with a great dose of empathy and recognition of the value of inter-dependence and limits of independence. Societies which teach empathy to children reach higher levels of trust managing more harmonious societies which produce more well-being to all. Companies which trust their workers and treat them with respect are more productive. The management of these forces has to be learned by experience. No amount of teaching by others will significantly increase our empathy and trust. Experience is the most effective teacher. Educational systems need to create the opportunities for such experiences to emerge rather than overwhelm students with repeated written wisdom in closed classrooms.

Examples of positive social synergy are plentiful and are given through most pages of this book. But examples of negative social synergy also exist. Here just a few:

**Social Parasitism** distributes wealth by taking away from the one who has, with no encouragement to increase global production, leading to general poverty. That is the popular base of Communism, Marxism, Populism and Nationalism.

**Economic Capitalistic Narcissism** is to claim the benefits of production, ignoring the agents involved in the processes that allow the emergence of synergistic interactions with others in the generation of profits.

**Wild Capitalism** is the consequence of practicing Economic Narcissism in an environment without Social Regulation.

**Resentment and Spite** are sentiments that seek destruction in order to achieve some sort of social compensation against perceived injustices.

The fundamental social cell where social synergy might emerge is the family, human work teams and human ecosystems. These cells aim to achieve:

- Psychological and physical safety for all individuals.
- Speeding up the accomplishment of needed tasks.
- Produce innovations.
- Maximize the acquisition and use of knowledge.

The working of group units has several limitations, Therefore, individual actions are favored over group action in many circumstances such as when the task to resolve is simple and does not require much innovation. Other problems with group work include:

- A group awakens the power needs of some of its members, causing authoritarianism, paternalism, maternalism and other behaviors that limit the action of the productive and innovative individuals to

emerge

- In a group, it is frequent that the one who speaks the most is not the one who can contribute the most to the common aims.
- Much energy and time is spend in communication and coordination
- The group dynamics limit individual innovation since group coordination consumes a lot of energy
- The team work is limited as its harmonious advance depends on the speed of the slowest of the group.
- The teamwork is inefficient if the coordinator or group leader has little knowledge about the tasks to be carried out
- Groups may foment authoritarianism. Limiting autonomous responsibility from each individual.

The aim of an optimal social contract is to achieve a productive ecosystem that harnesses the energy of different motivations, cultures and personalities without unnecessary expenses in bureaucratic structures.

### 5.7 Zero Sum and Win-Win Games

Extensive evidence shows that the evolution of human behavior occurred by different phases or steps [41]. Hunter gatherers lived first in small families. Later in evolutionary history, families organized into larger assemblies that forced the group to develop more sophisticated social rules. These Tribes or Clans were mainly linked through kin relationships. Social structures had an effect on the economic efficiency of the groups, so that clans associate with other clans or grew by incorporating non kin as members. In this phase, and later on, the survival of the groups depended mainly on the efficiencies of social synergies that could be achieved. Even today, the economic efficiency of European nations rest on a combination of features, grouped into a “property syndrome” [42] that maximizes synergy. The lack or poor development of one of these features reduces synergy enormously.

The features that allow citizens in a modern prosperous society to interact synergistically are trust and knowledge upheld by institutions such as law and order and universities. These institutions are not easy to establish and maintain as the collapse of Afghanistan, Syria, Venezuela and others has dramatically evidenced.

The main impediment in making the transition from hunter gatherer to modern state is the fact that optimal behaviors for zero sum interactions differ from those of win-win games.

An example Is the advice of the Bible (Deuteronomy 20:10-20). It describes the working of an instinct, common to most animals when confronted with competitors of the same species: “Put to the sword all the men in it”, and for the women and children, “you may take these as plunder for yourselves” and reproduce. This advice is still followed by many human cultures today. It is a zero sum game that minimizes the reproductive success of competitors and maximizes the fitness of the actor. Another example is the hunting of a deer in a forest. Once a hunter has killed it, it cannot be hunted by another hunter. Cooperative hunting, however, allows a group of hunters to capture a mammoth or a giant whale, which a single hunter is unable to capture. Here cooperation achieves synergy to the benefit of all involved.

Organisms optimize zero sum games often simultaneously with cooperation, producing a wonderful rainbow of behaviors. A schematic representation of the different forms of social interaction and their effect on the individual and the society, based on the results of the simulations with Sociodynamica is presented in Figure 5.4. The figure illustrates that in their overall effect on society, parasitic interactions and the pure altruistic actions are detrimental. They both represent interactions of “zero sum”. That is, what one loses, the other wins and vice versa. The qualifier altruistic or parasitic only depends on the point of view of the agent who suffers the interaction, and in both cases, the result is negative on the aggregate. It is only with the presence of some synergistic effect, which creates value, that interaction can increase the aggregate utility of the system. In this case, the interaction is not zero sum and both agents win. It is the so- called “win-win” interaction. These are the interactions that produce wealth. In real life, most useful interactions or exchanges are rather dissipative, since the transfer of utility or wealth, by fulfilling the second law of thermodynamics in a closed system, produces losses. That is, interactions that do not produce any benefit to the recipient and cost energy or time to the actor. These interactions lower the aggregate wealth of the system since they waste or dissipate it.

**Four basic types of social interactions**

| Action | Effect | Behavior |
|---|---|---|
| **Wise** | Both the individual and society win | Social investment |
| **Selfish** | Individual wins, society looses | Destructive selfishness |
| **Altruistic** | Society wins, individual looses | True altruism |
| **Stupid** | Individual and society loose | Destructive Behaviors |

In short, the diversity of actions and their effects on society can be classified into four broad categories as listed in the Table

Now then, the interactions we humans carry out have been selected and decanted by biological evolution and our cultural history. Consequently, they must have an adaptive reason. Purely dissipative interactions can be considered non-adaptive and clearly all societies reject them. They are considered as social aberrations, antisocial conducts or negative behaviors.

Interactions of “zero sum”, especially the selfish ones, are the type that has dominated the social interactions of *H. sapiens* for most of its history. It is the interaction of the hunter-gatherer with non-genetically related peers. The prey or fruit that an individual donates or takes is not available to others. The fish he extracts from the river cannot be caught by another fisherman. These limitations of the environment favor mutualistic interactions in which every act of generosity is expected to be rewarded in the future. It is the basis of economic ideologies that prioritize the distribution of wealth. This way of seeing the world can be summarized as “God creates the resources, we humans hand them out.”

On the other hand, synergistic interactions have very powerful properties and the actions that promote them can be considered as an investment in the short and/or long term, as they are creators of wealth. Its omnipotent presence is relatively recent in human societies. Its operation is based on technology and knowledge. The creation of technological empires is not possible by the simple addition of wills. It also requires the complement of knowledge and skills, the enhancement of cooperative interaction with a broad understanding of sciences and appropriate technologies. It is a product of the scientific and industrial revolution. The optimized cooperation between workers, technicians, managers and financiers in a technology company, creates an added value that is much greater than that created by the interaction among the same number of subsistence farmers.

Another example that illustrates the working of synergy: A typical top team and the people who follow its activities focus on business issues and whether the team got the strategies right. Little attention is paid to the way members of the team interact personally to propose, develop, and apply these strategies. Formal team-building events, behavioral interventions, and facilitated workshops to improve a team's performance are rare. Business-performance issues get direct and immediate attention; behavioral issues are addressed, if at all, indirectly and after the fact. Yet top teams don't come together by magic. Improving their performance as teams should be an explicit goal realized through explicit means. (McKinsey Classics |

July 2021)

### 5.8 How to limit parasitic governments

Lessons from the evolution of social parasites [43].

During the process of coevolution, social parasites have evolved sophisticated strategies to exploit their social hosts. Slave-making ant queens invade host colonies and kill or eject all adult host ants. Host workers, which enclose from the remaining brood, are tricked into caring for the parasite brood. In following raids stolen host broods are similarly enslaved. Evolution of anti-parasite adaptations also exist in the form of rebellion of enslaved *Temnothorax* workers, which kill two-thirds of the female pupae of the slave-making ant *Protomognathus americanus*. Thereby the co-evolutionary battle between host and social parasite may be won sometimes by the host. Other examples include hosts of avian brood parasites, where perfect mimicry of parasite eggs leads to the evolution of chick recognition as a second line of defense.

Similarly, governments in human evolution co-evolved with their societies in a dynamic balance where sometimes governors become parasites of their society (autocrats, dictators, live-long governors); i.e., profit more from their social symbiosis with their hosts than the general population. Democracy has been purposely evolved to counter such swings in the governor - host balance that may severely harm large sectors of human populations. Yet often restricted or false information, and manipulation of facilities guiding elections, can and are manipulated to favor governors in detriment to the governed. Ancient Greek and Venetian political scientists developed “demarchy” to prevent these shortcomings of democracy. Other evolutionary developments of democracy include the creation of institutions ( “independent” judiciary system and Central Banks) as instances to avoid politicization; i.e. grab of power by interest groups from those with less access to information and to governors. Demarchy makes it more difficult for these institutions to loose their independence. Demarchy implements democracy by randomly selecting electors that will have direct access to relevant information and to those being elected. Modern applications of demarchy [44] include the selection of Judges, of members to the Supreme Court, of directors of the Electoral College, and making fundamental constitutional decisions. Here, a grand jury or sworn body of people (the jurors) convenes to render a verdict officially submitted to them. Jurors are selected at random from a representative sample of the population and are thus more likely to be independent from the power of governors.

A vulnerability of parasites in general is that they lose the capacity for some fundamental processes and thus become dependent on their host.

## 5.9 Bioeconomics and Sociophysiology

The insight gained in recent years by biologists and economists into the working of evolution of complex systems is impressive. Unfortunately, little communication flows between these two academic guilds. Both guilds recognize that synergies are the drivers of evolution. Thus, it seems worthwhile to attempt a synthetic framework that encompasses both approaches. A branch of bioeconomics is doing just that.

In biology, William D. Hamilton's (born 1936 in Cairo and died 2000 in London) original Inclusive Fitness Theory[45], later expanded by David Queller[46] (born 1954, Madison, Wi), explains the conditions that favor the emergence and maintenance of social cooperation by synergies achieved through social cooperation. In economics, a similar phenomenon occurs[47]. That is, evolution will favor any behavior of an individual that increases its benefits and its likelihood to have more children. The novelty here is that behaviors that favor others, but that will trigger indirect actions that eventually benefit the initiator of the behavior, are treated by evolution in the same way as egoistic behaviors. And if synergy increases the benefits to all actors, evolution will favor those behaviors even more. This mechanism guarantees that behaviors that favor social long term social investments are feasible in evolutionary terms and are evolutionary stable.

The conditions that favor the establishment of social long-term investment through synergistic cooperation are, in addition to the level of synergy involved, kin-relationships and assortation or homophily that joins like with like. That is, individuals engaging in synergistic cooperation that share similar genes, behaviors, tastes, motivations, aims, communication systems or other devices that favor cooperation, are more likely to be favored by adaptation, or the selection processes working in evolution.

In bioeconomic terms, the theory that explains the dynamics of long term social investments aided by social synergy is called the Extended Inclusive Fitness Theory (EIFT)[48] which synthesizes the natural selection forces acting on biological evolution and on human economic interactions by assuming that natural selection driven by inclusive fitness produces agents with utility functions that exploit assortation and synergistic opportunities. This formulation allows to estimate sustainable cost/benefit threshold ratios of cooperation among organisms and/or economic agents, using existent analytical tools, illuminating our understanding of the dynamic nature of society, the evolution of cooperation among kin and non-kin, inter-specific cooperation, coevolution, symbioses, division of labor and social synergies. EIFT helps to promote an interdisciplinary cross fertilization of the understanding of synergy by, for example, allowing to describe the role for division of labor in the emergence of social synergies, providing an integrated framework for the study of both, biological evolution of social behavior and economic market dynamics.

### *5.10 Bioeconomic Utility Function*

These evolutionary forces develop behaviors and motivations that optimize the cooperation mechanisms. These include everything related to sex, power, affection, reputation, resentment, shame, guilt and many other elements. These social feelings are fundamental for the decision making of each individual and must be included in their utility function.

The utility function is a function that measures the "satisfaction" or "utility" that a consumer obtains when enjoying goods and determines the particular economic behavior of each person. In bioeconomy this function is defined not only for the behavior of consumption of goods and services but for any decision that a person can make during his life. This expanded definition for utility function explains much better the economic and non-economic behavior of humans

An extreme example is a bio-economic understanding of the motivations of terrorists, which identifies different forms of terrorism. If a person, due to some strange quirks in the establishment of their utility function, associates infinite benefits with immortality, or nil chances of success due to a perceived irredeemable fatality, strange behaviors appear. Terrorists optimize features in their utility function that are seen as pathological by others but are rational in terms of their utility function. Understanding the biological roots of utility functions in humans will improve our understanding of the motivations of normal and abnormal human behaviors, including that of terrorists.

Another particular approach of bioeconomics is sociophysiology. This approach draws upon tools of physiological sciences to understand social interactions, and vice versa). A seminal example is that provided by Keiichi Masuko (born 1954 in Tokyo, Japan) using ant colonies. He showed that larvae workers may feed the mother queen even with their own blood. Thus, the flux of energy and matter inside a colony of organisms can be viewed, at the same time, as an economic dynamic and a physiological process. The insights from physiology to economics are that rather than analyzing economic phenomena from the perspective of equilibrium, even of dynamic equilibrium, the concept of homeostasis seems to be more appropriate. This more broad concept fits better with that of stable states of the physical chemist Ilia Prigogine. These concepts seem much better to describe complex economic dynamics that operate far from equilibrium and where irreversible processes participate. These insights open the door to other alternative views that have much higher explanatory power than linear equilibrium models, commonly used in reversible thermodynamics and in classic economics.

## 6. Alternative representations of Synergy

The parable of the blind men and an elephant from ancient India tells of a group of blind men who have never come across an elephant before, learn and conceptualize what the elephant is like by touching it. Each blind man feels a different part of the elephant body, but only one part, such as the side or the tusk. They then describe the elephant based on their partial experience and their descriptions are in complete disagreement on what an elephant is. The natural elements behind what we call synergy is like a nondescript “elephant”. Several divergent descriptions of it exist. Among them, are the concept “emergence”[49], self-organization[50], hierarchical organization[51], thermodynamic complexity[52], stationary open systems[53], and the Constructal law[54].

### *6.1 Emergence*

In philosophy, systems theory, science, and art, **emergence** is a phenomenon whereby larger entities arise through interactions among smaller or simpler entities such that the larger entities exhibit properties the smaller/simpler entities do not exhibit.

Emergence is central in theories of integrative levels and of complex systems. For instance, the phenomenon of *life* as studied in biology is an emergent property of chemistry, and psychological phenomena emerge from the neurobiological phenomena of living things.

The concept “emergence” looks at the process of increasing some kind of order and of work efficiency – i.e. synergy - from the perspective of an ingenious observer, where properties emerge that are beyond the grasp of the observer. The concept focuses on the mysterious aspects of this process, whereas synergy focuses more on the phenomenological and mechanistic aspects of this process.

The concept emergence is intuitively very potent, but only reflects our ignorance about the underlying mechanisms responsible for the phenomenon. For example, the emergence of the properties of water, when oxygen and hydrogen atoms are made to react chemically, seem to be mystical. Quantum mechanics explains this process, erasing the mysticism out of this “emergence”.

### *6.2 Self-Organization*

Ilya Prigogine (1919, Moscow, 2003, Brussels) and Jean-Louis Deneubourg (1951, Ath, Belgium) promoted the concept of **Self-organization**, also called spontaneous order (in the social sciences), is a process where some form of overall order arises from local interactions between parts of an initially disordered system. The process is spontaneous, not needing control by any external agent. It is often triggered by random fluctuations, amplified by positive feedback. The resulting organization is wholly decentralized, distributed over all the components of the system. As such, the organization is typically robust and able to survive or self-repair substantial perturbation. For example, chaos theory discusses self-organization in terms of islands of predictability in a sea of chaotic unpredictability.

Self-organization occurs in many physical, chemical, biological, robotic, and cognitive systems. Examples can be found in crystallization [55], thermal convection of fluids, chemical oscillation, animal swarming, and artificial and biological neural networks. The concept self-organization may be applied to synergistic and non-synergistic processes (Process Type II and III in Table 6.1). For example, self-organization of particles settling as sediments in a pond, or atoms settling in a crystal are self-organized processes but are not synergistic. Although entropy of the system decreases, so does the free energy or potential work that it can perform. Self-organization focuses on the fact that some processes are not directed from outside the system by a human or other entity. It has thus an anthropocentric taste to it. Synergistic self-organized processes, as those described in this book, are organized by physical and thermodynamic laws. This fact is diluted by referring to this process as self-organized.

## *6.3 Constructal Law*

Adrian Bejan (Born 1948 Galati, Romania) proposed in 1996 a thermodynamic law for all design generation and evolution phenomena in nature, bio and non-bio. The **constructal law** represents three steps toward making "design in nature" a concept and law-based domain in science:

1. Life is flow: all flow systems are living systems, the

animate and the inanimate.

2. Design generation and evolution is a phenomenon of physics.

3. Designs have the universal tendency to evolve in a certain direction in time.

The constructal law is theory in physics concerning the generation of design (configurations, patterns, geometry) in nature. It holds that shape and structure arise to facilitate flow. The designs that happen spontaneously in nature reflect this tendency: they allow entities to flow more easily – to measurably move more current farther and faster per unit of useful energy consumed. Rain drops, for example, coalesce and move together, generating rivulets, streams and the mighty river basins of the world because this design allows them to move more easily.

The constructal law asks the question: Why does this design arise at all? Why can't the water just seep through the ground? The constructal law provides this answer: Because the water flows better with design. The constructal law covers the tendency of nature to generate designs to facilitate flow.

A recent definition of this law states that for a finite-size system to persist in time (to live), it must evolve in such a way that it provides easier access to the imposed currents that flow through it. It describes the natural tendency of flow systems (e.g. rivers, trees and branches, lungs, tectonic plates and engineered forms) to generate and evolve structures that increase flow access

The Constructal law focuses on the flow design that changes in a discernible direction over time, toward easier flow access, and 'constructal', which comes from *construere* in Latin. Synergy focuses on the mechanisms powering this process analyzed in a given moment of time. Like in a movie that consists of multiple framed images, the Constructal law is a kind of Synergmatics that focuses on the moving pictures that reveal synergy. Synergmatics requires the understanding of Synergy. Thus, synergy must be the foundation of any constructal analysis.

# 7. History of Synergy

The concept of SYNERGY has a known history. We can pinpoint the emergence of the concept to three pioneers: Charles Scott Sherrington, Hermann Haken, and Richard Buckminster Fuller. Here a brief description of their relevant insights.

## 7.1 Coordination in Anatomical Physiology

In 1910, the physiologist Charles Scott Sherrington pointed out that the modular organization of the motor control system is based on the flexor reflex as the mother of all modules and synergies, He proposed that stepping was basically a series of flexion reflexes, with extension occurring merely as the “rebound” following the flexion. The extension during the stance phase of gait could be provided as some type of “extensor thrust,” evoked by “the weight of the animal applied through the foot against the ground”. His experimental studies on reciprocal innervation of muscles showed that the relationship between the effort made as input and the work output was nonlinear and had synergistic properties, winning him the Nobel Prize in Physiology and Medicine in 1932. Today we know that the complex way muscles, tendons, nerves and bones interact produces mechanical efficiencies that are superior to what mechanical engineers have achieved so far. Thus, plenty of additional synergies in neuron-muscular mechanics remain to be discovered.

The complementarity of the two brain hemispheres (left and right) in mammals, and specially in humans, also elicit synergistic properties [56]. Each hemisphere complements the functions of the other and we need both to be full humans and act normally. That is also the case of our two hands (left and right) which complement synergistically their capacities, making a one handed human, irrespective of which hand is missing, to strongly diminish its manual capacities [57].

### 7.2 Laser and strength of Coherence

Hermann Haken, a pioneer of irreversible thermodynamics developed an interpretation of the laser principles as self-organization of non-equilibrium systems paved the way at the end of the 1960s to the development of synergistics.

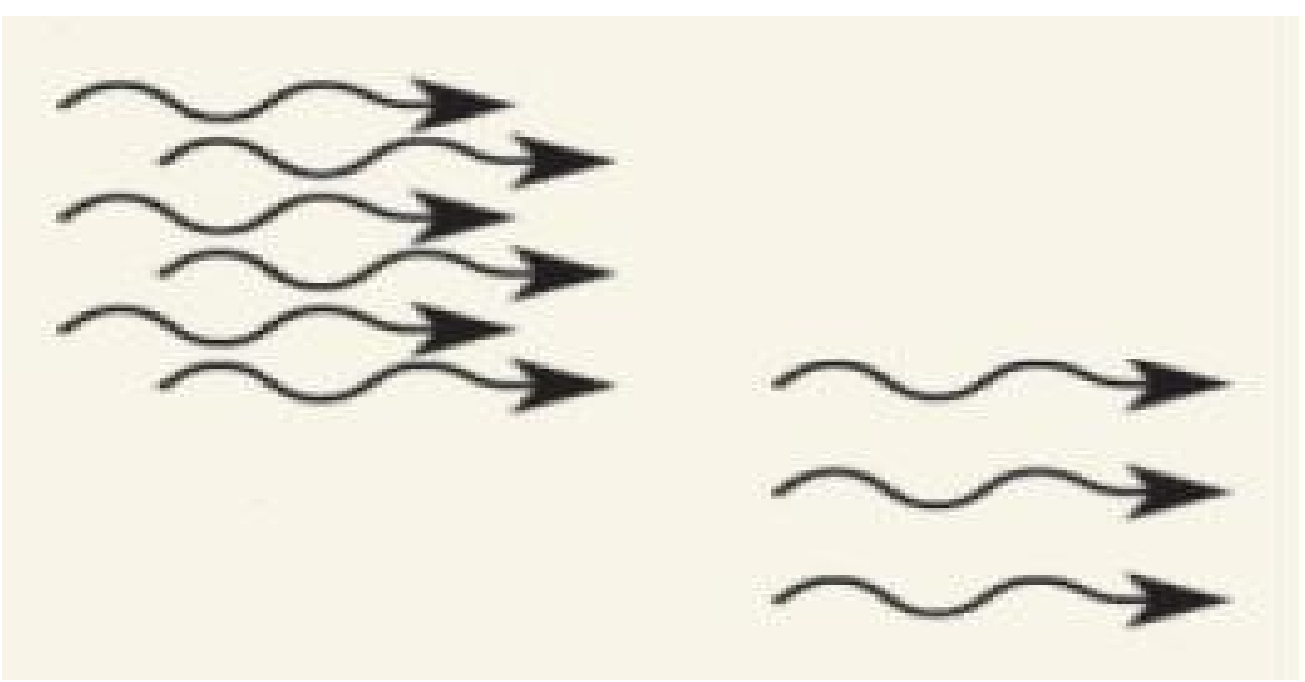

**Figure 7.2**

A few laser light rays (bottom right) are much more potent than many rays of non-coherent light (top left). Source: drawing by the author adapted from Google Images

He explained the production of laser light, highly ordered light or coherent light (three undulating arrows at the bottom right of the figure) instead of noisy light (undulating arrows in the upper left corner of the figure) as a nonlinear phase transition when enclosed gas atoms are activated with an undulating electric field. When increasing the strength of the electric field (order parameter), more light is produced, but only after a certain threshold is surpassed, does the gas atoms engage in stimulated emission of coherent light.

In coherent light (bottom right of the figure) or laser light, each light ray adds more to the final light intensity than in non-coherent light (top left of the figure). Thus, a light beam with few rays of laser light is much more potent than a beam formed by many non-coherent rays of light. In this sense, the addition of laser rays is synergistic in relation to the addition of non-coherent rays of normal light.

This principle can be applied to many situations where a common goal is achievable. That is, synergy can be achieved in many different physical domains by coordinating diverse types of forces. Thus, for each example, the focus of the synergistic process has to be identified. Haken presents a large number of very diverse examples which can be reviewed in his book “The Science of Structure: Synergetics”[58]

### *7.3 Architectural Synergy*

Synergy as parts supporting a whole. The term synergy was refined by Richard Buckminster Fuller (born

1895 in Milton, MA and died 1983 in Los Angeles) who analyzed some of its implications in architecture coining the term Synergetics. In his book “Synergetics: Explorations in the geometry of Thinking" published in 1975, he refers to this process as a mystical phenomenon, describing it as a behavior of integral, aggregate, whole systems unpredictable behaviors of any of their components or subassemblies of their components taken separately from the whole.

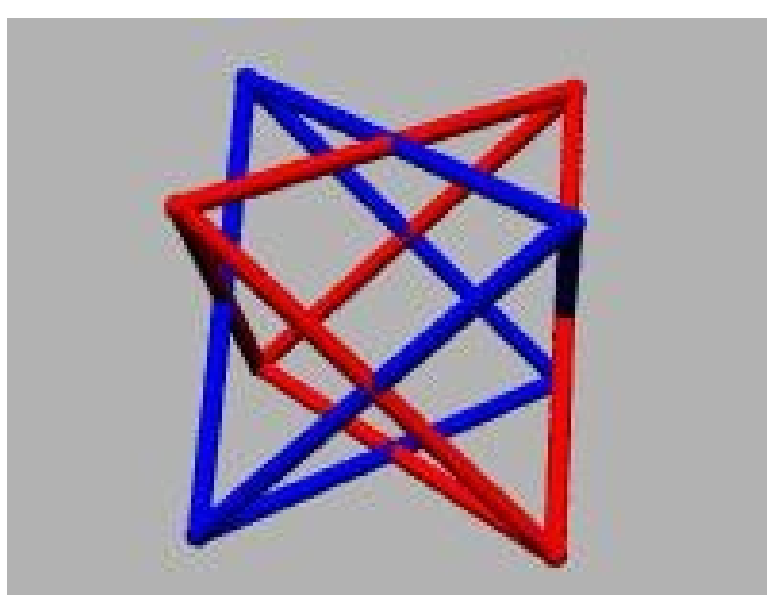

**Figure 7.3**

Robust architectural structures: Source: Google Images

His discovery arose from the fact that the mechanical resilience, robustness and carrying capacity of a physical structure depends very much on the way its parts interact. This was already known in biology (see example 3.1) but not at his time applied in engineering and architecture. For example, the structures he designed, such as those in the Figure are much stronger than other structures using the same sub-components but arranged in other ways. This led him to design the geodesic dome which made him famous worldwide. The structures supporting the domes were much lighter, robust, beautiful and cheaper than the classical alternatives. Now we know that nature had invented this structure thousands of millions of years before and created "Fullerenes" or carbon structures using this design.

# 8. Measuring Synergy

## *8.1 Mathematical description for Synergy*

(For analytical minded readers only)

No unambiguous definition exists for synergy but several non-additive phenomena might qualify as synergy. In the world of many components or constituent parts, complex interactions are not always arithmetically conserved. That is, two forces may add up to less than what the sum of each one would yield when measured separately, because they dissipate energy when joined; they might conserve their energy perfectly and sum up arithmetically their energies; or they might interact synergistically releasing more energy that what they both represent when measured separately. We will describe these different ways of interactions with a scalar that we call the synergy constant of interactions or Φ in Greek letters. Thus defined, the value of Φ would be less than 1 in the first case where energies are dissipated when forces are joined; Φ would be equal to 1 in the case of ideal conservative interactions, and the value of Φ would be larger than 1 when synergies between the two forces are released.

An empirical observation shows that all known synergistic processes studied so far [59], where $\Phi > 1$, are characterized by a decrease in entropy or, what is the same, an increase in negentropy (N), coupled to an increase in free energy available or the potential useful work that the system can perform (W). As N and W are measured in a variety of different units that are not easily translated into each other, we calculate a scalar that is independent of the unit of measurement This scalar $\Omega$ is defined as the proportion of change before and after the synergistic process occurred. This proportion of change can be calculated for $\Phi$, N, and W. Empirical evidence shows that for processes widely recognized as synergistic, the following proportions holds:

$\Omega(\Phi) > 1$, $\Omega(N) > 1$, $W > 1$

Thus, any feature in a system that is amenable to measurement and that is somehow related to the structure of the system might serve to estimate negentropy. And any measurable effect of actions of the system that reflect its strength or efficiency allows to build an index that estimates useful work. These indices are then used to calculate unit-less ratios of change. These ratios of change can then be used to assess if synergy is involved in the process we want to study. If both these ratios are positive, we can be sure that we are dealing with synergy. If both these ratios are negative, we are sure to be in the presence of a Carnot type process analogous to combustion. An undetermined number of intermediate possibilities exist that will provide continuous fodder for future research (see Table 8.1).

**Table 8.1:** Types of thermodynamic processes

| **Type** | **Name** | **Free Energy** | **Entropy** | **Neg-entropy** | **Process** | **Example** |
|---|---|---|---|---|---|---|
| I | Dissipation | - | + | - | Combustion | Engine |
| II | Synergy | + | - | + | Life | Organism |
| III | Settling to Equilibrium | - | - | + | Crystallization | Salt Solution |
| IV | Disordered Growth | + | + | - | War<br>Stock-Market | Winning army<br>Average Investor |

The Table characterizes four different types of thermodynamic processes that vary in the way that increases or decreases free energy and entropy. It shows that only in synergistic processes, free energy increases and entropy decreases.

Defining very large systems, like the winning party in a war, is problematic. The system changes during the process, and subsystems in that system might suffer dissipative or synergistic processes. For example, sectors of the winning party might suffer heavy destruction, whereas other sectors might develop a highly sophisticated organization.

Processes might change over time and the type of process detected depends on the time window used. In the case of an investor, an increased instability after a large investment might be temporary, until profits from the investment crystallize, then the process might become synergistic. This underlies the logic of the concept of "Creative Destruction" proposed by Joseph Schumpeter (Born: Triesch, Check Republic, Died: Taconic, Connecticut, USA). This process occurs in biological evolution through natural selection: the extinction of species allows new ones to prosper. In economic evolution, elimination of outdated and failed enterprises allows new and better ones to emerge, driving economic progress synergistically to new heights.

## 8.2 Quantifying Social Synergy

Watt / individual /hour

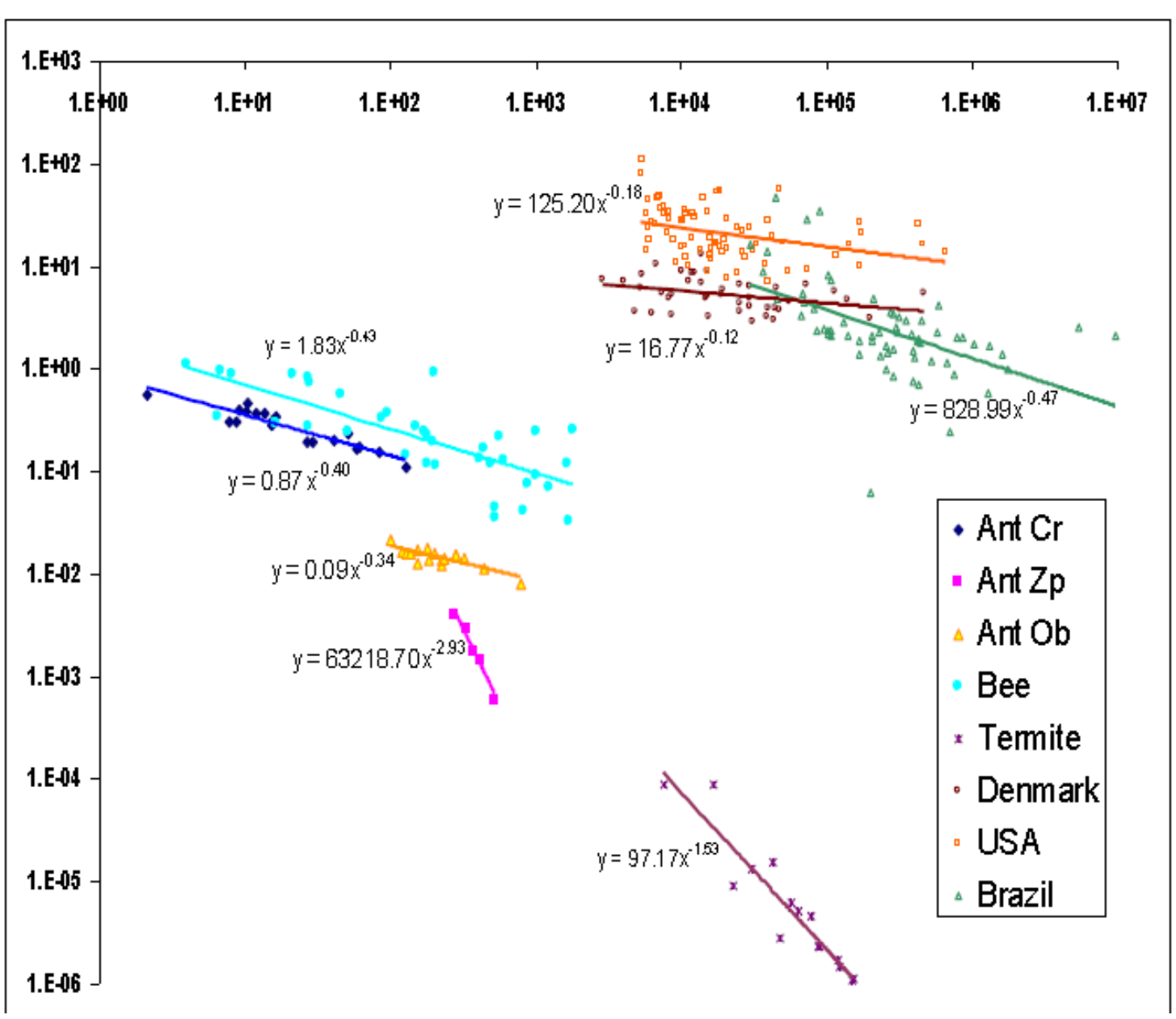

Number of individuals (log scale)

**Figure 8.2:** Logarithm of energy consumption in watts per individual per hour (vertical axis), against the number of individuals in the society (horizontal axis) for Ants (Cr: *Camponotus rufipes*, Zp: *Zacryptocerus pusillus*, Ob: *Odontomachus bauri*, Termites, and Human cities in Denmark, USA and Brazil. [60].

There are many ways to quantify Φ, N, and W. In the figure, I give a graphical empirical example. Here, the energy consumption of individuals living in societies of different sizes is measured. This measurement, be it in terms of oxygen consumption per individual, amount of carbon dioxide produced through respiration per individual, or electricity consumption per individual are transformed into a uniform unit such as watt (which are joules per second) per individual, plotted for groupings of different sizes. Data comes from social insects (termites, ants and bees) and social mammals (humans). Units for insects are ml of $CO_2$ per minute and those for human cities are KW of electricity consumption per year.

The plot shows that as social aggregates become larger, with more individuals in the society, the energy consumption per individual is reduced. That means that individuals living in larger societies need less energy to survive than individuals living in smaller societies or living alone. The explanation for this difference is that society makes living more efficient. This increase in efficiency is proportional to the rate of reduction in energy consumption as societies become larger. Thus, an indicator of social synergy can be measured quantitatively.

This indicator allows us to compare different societies. In the examples given in the figure, the societies which achieved the largest social synergy among insects were *Zacryptocerus* ants and termites, and among human cities sampled, Brazil. Brazil had the largest income inequality among its citizens of the three countries studied and *Zacryptocerus* ants and termites the largest variance in morphology (polymorphism) among its workers. As larger societies consume more total energy than smaller ones, they are likely to produce more total useful work in aggregate. But the data per capita show that in addition, we can relate synergy with complexity or negentropy and useful work or free energy in thermodynamic terms.

Other complex systems are also amenable to quantitative measurements of synergy. Comparing the average literary text with those written by Nobel Prize laureates shows a difference in complexity that allows us to calculate the synergy required to explain the difference. Similar calculations can be made comparing old and newer classical music, languages, computer codes or brain complexity as related to social complexity. In all cases, synergy, complexity or negentropy, and the potential for producing usable work can be calculated.

## 9. How to Achieve Synergy

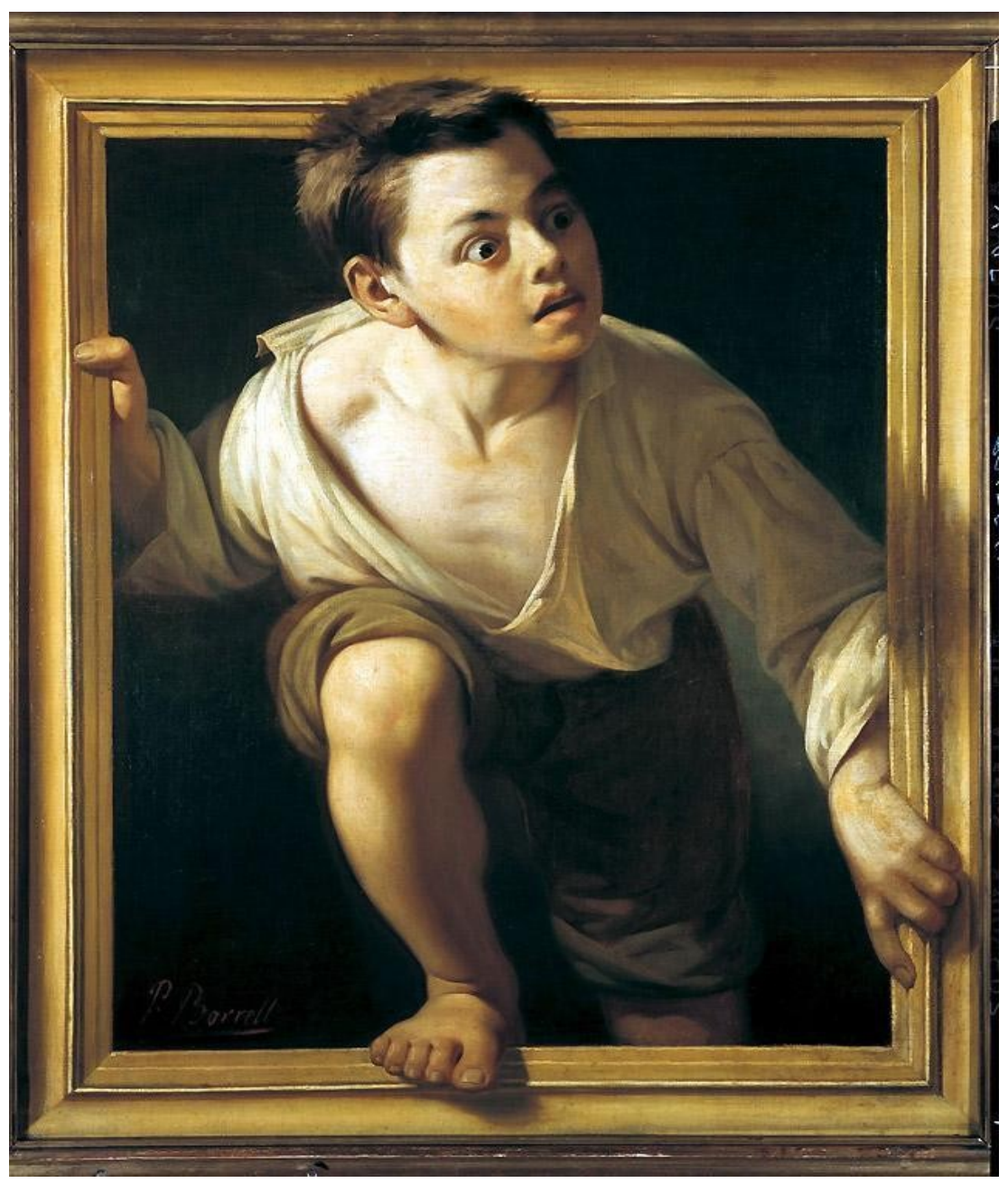

**Figure 9**

Synergy is often to be found in a new dimension. We need to step out from our traditional frame of mind to detect novelty, look and think out of the box. *“Escaping Criticism painted 1874” by Pere Borrell del Caso (1835 Puigcerdá, 1910 Barcelona)*

Distilling all the information about synergy we have extracted from nature, we can now identify the elements that allow for synergistic processes to occur, or to identify the roots of synergy. We know the fundamental component of any synergistic process, and might list some general practical recommendations as follows:

### 9.1 Random or Directed Variation

No novelty can emerge without diversity and freedom. Our mind is not capable to grasp the fine details of complex interactions and thus is badly prepared to predict precisely the abilities, skills and forms required to build synergies to perform a given task. Trial and error are the best guide to achieve this. But for trial and error to work, we need diversity, variance in the inputs, and continuous innovation and novel proposals.

Different kinds of diversity produce different environments and favor different industries. Regions with hundreds of different types of cheese have a different industry to that of regions with a large diversity of beers. That is, there is a reciprocal feedback between diversity and economic activity. Each one affects the other. Its management must be very subtle, nudging rather than planning and evaluating continuously the effect of different actions. Random mutations, errors and mishaps are part of the process. They must be taken account of and accepted.

The ease with which bankruptcies are handled in the US is a cause of the success of its economy. It favors inventiveness and promotes innovative motivations. Handling the right balance between creativity and innovation and productive discipline and financial rigor. Each situation must be evaluated on its own merits and the balance must be adapted according to the results that are being obtained.

## 9.2 Exploit Division of Labor

Once we have a diverse group in an innovative environment, spontaneous selection guided by clear benchmarks of productivity or efficiency will help to spontaneously auto-organize the system to produce emergent elements based on specialization of the participating parts. To guarantee the desired outcome from this dynamic, we need to guarantee 6 axes of action.

### 1. Establish clear benchmarks

To guarantee progress, you need to know if you or your project are advancing or receding. To know if your ship or train is moving, you must fix a reference point and watch if your relative position to it is changing. Without quantifiable and easily verifiable benchmarks, your boat is at drift.

### 2. Look for the best

It is difficult to make good grape juice with rotten grapes, unless you purposely ferment the grapes properly because you want to produce wine. Regarding quality, there is no linear relationship between no quality and top quality. The best is always unbeatable. That is the definition of “The Best”. If you aim for success, never settle for inferior choices. Look for The Best and do your best!

### 3. Reduced negative interactions

Wars destroy economies, strife ruins partnerships, conflict bankrupt otherwise prosperous enterprises. The single most damaging feature of any dynamic system is conflict. Avoiding it is everyday homework. Conflicts can start from the least expected corner. Guaranteeing harmony always and everywhere is best to avoid conflict that dissipates past investments and costly efforts.

### 4. Search for Complementarity

The single most important predictor of synergy in a setting of agents or individuals dividing labor is complementarity. Stepping over your neighbor's toes doesn't help you nor your neighbor to advance. To complement each other’s action requires a holistic outlook that pinpoints the missing aspects of a whole. To complement each other requires respect for yourself and for the situation and action of others, and a deep understanding of the detailed dynamics of the enterprise. Technical and scientific knowledge are the best aid to achieve it and assortation joining like with like minimizes dissipation agonistic interactions between the parts.

### 5. Promote free choice in interactions

No individual is in possession of "The Absolute Truth", unless you are convinced that you are God. If so, your proper place is not in civilized human Society. In complex systems, local perspectives detect features and things that are invisible from other points of the system. Collective intelligence by far outsmarts the individual kind. But even actors in a specific point in a complex system might not spell out the details they see. Let them decide local issues, they are the best placed to take fast, accurate decisions if the overall goal of the enterprise is clear.

### 6. Match motivation with responsibility

The best pianist is a person who likes playing piano. You will never excel in an activity you don't like. Motivation is the single most important root for success. Only doing what you love and for something or somebody you love will lead to excellence. If an individual feels to be an integral part of the group he/she will defend it and work for its improvement. If he/she feels marginalized and excluded, expect only the minimum of work and effort. With such poor motivation do not expect to achieve synergy

## 9.3 Maintain Continuous Adaptation

### 1. Synergy only emerges in open systems

Closed systems in equilibrium do not permit the emergence of synergy. Continuous innovation, flux of ideas, people and resources are required to form stable states that are the nests where synergy breeds. But take care: there are limits to complexity, innovation and creativity, and thus, to synergy. Too much diversity and too little coordination, just produces noise. Assortation minimizes this danger. If you want to sustain an activity, business or enterprise you need sustained involvement and management. Motivation is dynamic and requires continuous reinforcement for its maintenance. Matching the motivations and characters of people helps in producing innovation with little noise. The precise balance though must be worked out for each situation.

### 2. Synchrony: access to max information

Complex division of labor and activity requires continuous synchronization. As complex systems are flexible and variable, synchronizations are not easily programmable. Thus, continuous communication between the interacting parts is essential.

### 3. Dynamism

Excellence requires continuous adaptation. Adaptation requires continuous change. A static system in equilibrium is dead. Life implies Change. Only by accepting, promoting and supervising continuous change can a complex system be kept alive.

### 4. Evaluation and selection

To maintain continuous adaptation of a system you need to evaluate its performance against ever adjusted and updated benchmarks, produce continuously better ways of doing things, test these alternatives and select the best. We know that synergy has been achieved when the system produces large increases in work capacity and large increases in the use of non-redundant information.

## 5. Do not stop initiative, direct them

You never know where and when the next best solution is going to emerge. If somebody has thought of a solution and makes a proposal, use this effort and motivation for your goals. Do not say NO, it cuts motivation and thus creation. Say YES and challenge the proposer to adapt the proposition to meet certain goals. Make everybody be part of the team and make them feel it.

## 9.4 Apply Empirical Rationality.

Acquired and accumulated information must be transmitted to others. Heredity, learning and teaching is thus a fundamental part of all successful evolutionary processes. The kind of information transmuted must be carefully selected. But accumulated information by itself is insufficient to maintain synergies. It has to be updated continuously.

The human mind has severe limits in tackling complexity, so that our abilities to construct theories and other constructs of our mind are very limited. The most important human insights on nature are empirical, not theoretical. Here are some examples of empirical laws that have no theoretical explanation. In the Kantian sense, they are scientific *a priories* or fundamental natural laws upon which our scientific knowledge is built. Some of them are:

- Life is an irreversible process: Aging and death are empirical certainties.
- The first law of thermodynamics: energy is always conserved

- The second law of thermodynamics: all processes produce entropy.
- The third law of thermodynamics: entropy and temperature have absolute values.
- The laws of gravitation as discovered by Galileo and formalized by Newton
- The laws of relativity: the speed of light is constant in a vacuum.

Achievement of synergy is also an empirical experience. Many management theories come and go. Synergy is here to stay as it has its roots in thermodynamics. But no simple recommendation for achieving synergy can be made. As any complex process, it is rather a syndrome of several features that must be working smoothly to emerge.

An illustrative analogy is the working of a car. No single element makes a car run smoothly. We need petrol and battery, oil and air in the tires, a steering wheel and a front window. All the elements and many more are required for a car to work properly. As so it is with synergy. Many elements are required to work together to achieve it. Many theoretical recommendations will be given, but empirical experience will provide the last word.

Fundamental laws of the working of synergies have been unveiled, the social engineering part of its establishment must be worked out now.

## 9.5 An example: Synergistic Conversation

In a monologue the interaction is mainly unidirectional and the communicator has to maintain the attention of the public using tricks and tools that we normally associate with the arts. In a conversation, all participants interact. If after the conversation, all or at least one participant emerges with an improved and/or broader world view, we achieve a synergistic conversation. If your intention is solely to prove your ideas are right, you will nor have a synergistic conversation but a power tournament. If you want to excel in monologues, you might learn a lot from artists and stage performers. Power games profit from knowledge in ethology and psychology. Yet to achieve successful synergistic conversations other features and tricks are needed.

Some relevant features:

- Start finding a group of intelligent, knowledgeable diverse group of people
- Accept that you might be wrong or unknowable in certain fields and that others might know something you do not. Others have different experiences, use different time and space windows than you do, and pursue different interests. They have a different world view.

- Try to understand the world view of others and your place in that view so you can better explain your view.
- Pinpoint the relevant differences in these views looking for a level of analysis or a window in space and time that might bridge the gap between them.
- Build a synthetic world view that all can accept. If you succeed, you made a qualitative jump in knowledge.

Tricks to implement these features:

- Hear and listen first before you speak. Measure the time you speak with a watch. You might get excited and lose control of time, overwhelming others with a monologue. Remember: this is a dialog or a conversation among various independent minds.
- Never start an answer with “no”. Negative expressions from you will trigger defensive reactions in the listener, blocking rational thought and triggering feelings and automatic reactions.
- Ask questions to those that do not speak on their

own. You might be surprised to what they have to say.

- Do not offend. Offenses trigger rage which shut down rationality and unleash hate.
- Emotions are important. They tell others what you care for and increase their interest in these issues. But do not let emotions numb reason.

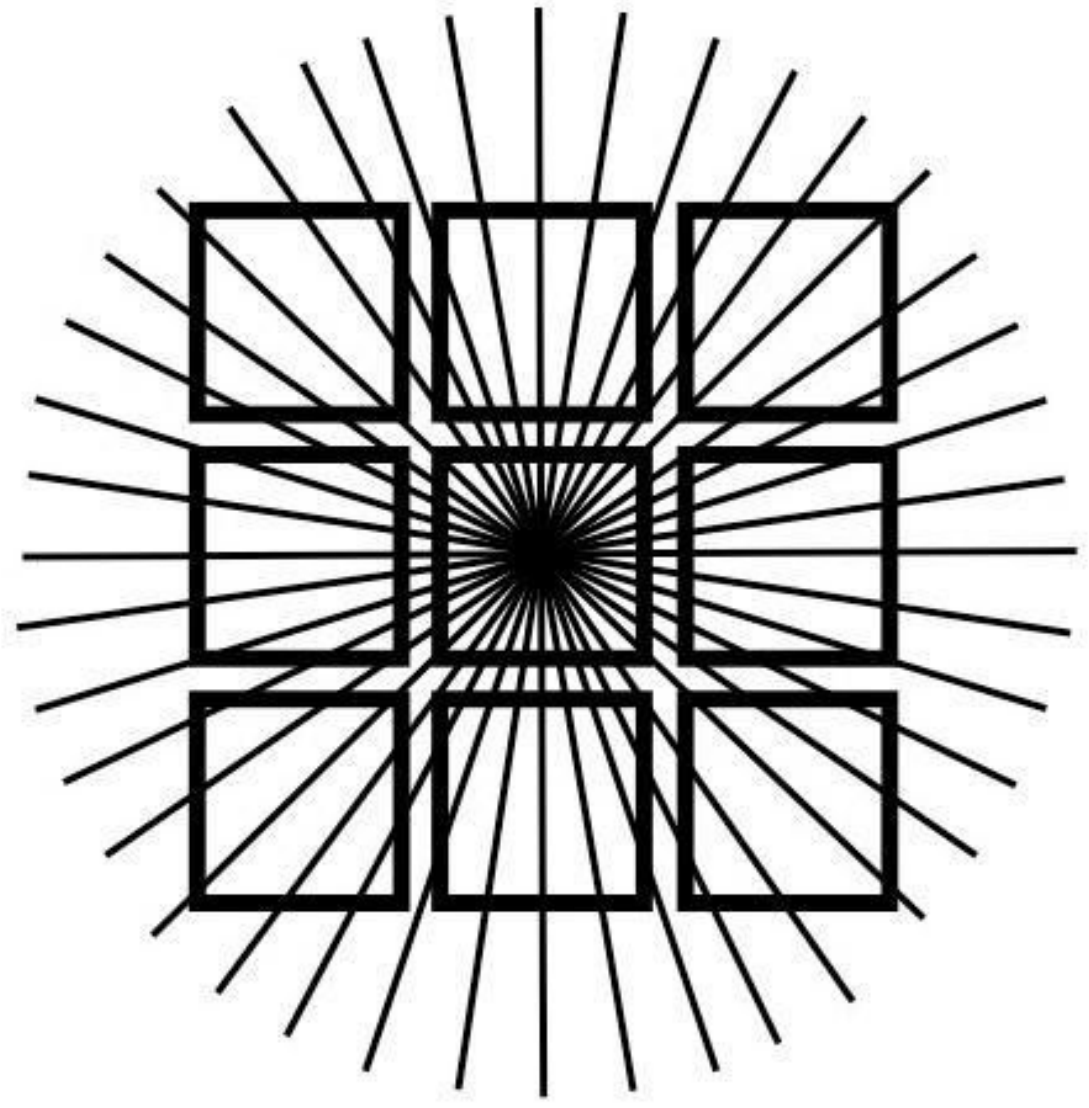

**Figure 10**

Our mind sees curves where there are straight lines and straight lines where there are curves. Empirical checks and experiments, as used in science, are the only useful criteria for validating theories and ideas. Source: Google Images

# 10. SUMMARY

Synergy arises from positive reciprocal and synchronized feedback loops in a network of diverse actors that exchange information, energy and matter. Its effect is to increase in a non-linear way (greater than the sum of the parts) useful work (free energy) and increase the order (decrease entropy) of the system. In contrast, noise is produced when information and incompatible actors are mixed. Greater free energy means greater emancipation from the environment, increased productivity, more efficiency, more adaptability, self-regulation and self-control of the system. Synergy arises from a synchronized division of labor with increasingly specialized.

After extensive exploration and analysis of examples of synergy in many areas of science, including physics, biology and economics, I postulate that the common elements in all these examples are essential for the existence of synergy. The fundamental elements or roots of synergy are:

1. Stable Open Systems
2. Cohesion
3. Diversity and Assortment
4. Complementarity
5. Communication and Synchrony
6. Freedom and Self-Organization
7. Dynamics with reduced entropy and free energy growth. This point is essential as it allows synergy to be measured and, therefore, the proposal is falsifiable and can be empirically refuted.

The actions to nurture synergy are:

**1 Promote diversity**

1. Import novelty, globalization, annealing, random and directed variations.
2. Maximize diversity and possible futures

**2 Exploit diversities through division of labor**

1. Establish individual benchmarks
2. Look for the best
3. Reduce negative interactions

4. Search for complementarity
5. Promote free choice in interactions
6. Match motivation with responsibility

**3 Maintain continuous adaptation**

1. Maintain your systems open and transparent
2. Synchronize actions with more information
3. Maintain dynamism
4. Evaluate and select continuously
5. Do not stop an initiative, direct it!

**4 Apply empirical rationality**

1. Practical results must guide theoretical thinking

**5 Look for the best**

1. In people, ideas, environment and everything you can

## Acknowledgments

Innumerable persons contributed ideas, discussions, criticism and knowledge during my life, which allowed me to crystallize the arguments and ideas presented in this book. I am unable to recall all of them. Thanks nonetheless to all. The manuscript profited from editorial reviews and suggestions from Jennifer Bernal, Gerardo Dias, Gerardo Febres, Andres Peña, Juan Carlos Correa, Adrian Bejan, Richard Cathcart, Rodolfo Jaffe, Yara Jaffe, Nancy Sajjadi, Claudia Lopez, Antonio Canova, Maria Corina Wallis Brandt, Cristina Lopez de Ceballos….

## Notes

[1] This book is available on Arxiv and Amazon under the name “The Roots of Synergy”, and at ELIVA under the name of “Information and Energy: Gods of Genesis”

[2] Jaffe K. “The Wealth of Nations: Complexity Science for an Interdisciplinary Approach in Economics”. Amazon books, 2014. ISBN-10: 1503252728

[3] See list in the Encyclopedia of Human Thermodynamics at www.eoht.info.

[4] Jaffe K. “What is Science? An interdisciplinary evolutionary view”. Amazon books, 2009. ISBN-10 1523271590

[5] Peter A. Corning. “Holistic Darwinism: Synergy, Cybernetics, and the Bioeconomics of Evolution, University of Chicago Press. 2005.

[6] Peter A Corning and Eörs Szathmáry. “Synergistic selection: A Darwinian frame for the evolution of complexity”. Journal of Theoretical Biology 371: 45–58, 2015.

[7] John Maynard Smith and Eörs Szathmáry. “The Major Transitions in Evolution”. Oxford University Press, Oxford, England. 1995.

[8] Van Valen L. 1973. A new evolutionary law.. Evolutionary Theory. 1: 1–30.

[9] McDonald MJ, Rice DP, Desai MM. 2016. Sex speeds adaptation by altering the dynamics of molecular evolution. Nature 531: 233–6.

[10] Jaffe K. Synergy from reproductive division of labor and genetic complexity drive the evolution of sex. 2018. J. Biol. Phys. 44: 317-320.

[11] Examples of the non validity of this ecaution hace been presented in Burgin, M. Diophantine and Non-Diophantine Aritmetics: Operations with Numbers in Science and Everyday Life, LANL, Preprint Mathematics GM/0108149, 2001, 27 p. (http://arXiv.org). Also in : WOLFRAM, Stephen. A Class of Models with the Potential to Represent Fundamental Physics (arXiv: 2004.08210). Y en: Jacob Lurie, Higher Topos Theory, 2009.

[12] Jaffe K. A Fourth Law of Thermodynamics: Synergy Increases Free Energy While Decreasing Entropy. https://www.preprints.org/manuscript/202007.0422/v1

[13] Relation between Constitutions, Socioeconomics and The Rule of Law: a quantitative thermodynamic approach. K Jaffe, E Martinez, AC Soares, JG Contreras, JC Correa, A Canova. http://arxiv.org/abs/2108.02094

[14] Two type of information: redundant and non-redundant heve been proposed by Rupert Riedl: “Die Strategie der Genesis. Naturgeschichte der realen Welt” 1976; o “Biology of Knowledge: The evolutionary basis of reason” 1984